\documentclass[10pt,twocolumn,compsoc,journal]{IEEEtran}%
\usepackage{graphicx,latexsym,subfigure,amsmath,epsfig,latexsym,subfigure,amsmath,cite,amssymb}
\usepackage{amsthm,mathrsfs,setspace}
\usepackage[ruled,linesnumbered]{algorithm2e}
\usepackage{multirow}
\usepackage{caption}

\usepackage{color}
\usepackage{bm}

\ifCLASSINFOpdf
\else
\fi

\begin{document}
\title{Two-dimensional Anti-jamming Mobile Communication Based on Reinforcement Learning}
{
\author{Liang Xiao,~\IEEEmembership{Senior Member,~IEEE,} Guoan Han, Donghua Jiang, Hongzi Zhu,~\IEEEmembership{Member,~IEEE,} \\Yanyong Zhang,~\IEEEmembership{Member,~IEEE,} and H. Vincent Poor,~\IEEEmembership{Fellow,~IEEE}\\
\IEEEcompsocitemizethanks{\IEEEcompsocthanksitem Liang Xiao, Guoan Han and Donghua Jiang  are with the Department of Communication Engineering, Xiamen University, China. E-mail: lxiao@xmu.edu.cn, hgajiayou@live.com, winky1508@outlook.com.
\IEEEcompsocthanksitem Hongzi Zhu is with the Department of Computer Science and Engineering, Shanghai Jiao Tong University, China. E-mail: hongzi@sjtu.edu.cn.
\IEEEcompsocthanksitem Yanyong Zhang is with the Wireless Information Networks Laboratory (WINLAB), Rutgers University, North Brunswick, USA. E-mail: yyzhang@winlab.rutgers.edu.
\IEEEcompsocthanksitem H.Vincent Poor is with the Department of Electrical Engineering, Princeton University, Princeton, NJ 08544. E-mail: poor@princeton.edu.}
\thanks{This work was supported in part by the U.S. National Science Foundation under Grants CMMI-1435778 and CNS-1456793.}
 }
}

\IEEEtitleabstractindextext{%
\begin{abstract}
By using smart radio devices, a jammer can dynamically change its jamming policy based on opposing security mechanisms; it can even induce the mobile device to enter a specific communication mode and then launch the jamming policy accordingly. On the other hand, mobile devices can exploit spread spectrum and user mobility to address both jamming and interference. In this paper, a two-dimensional anti-jamming mobile communication scheme is proposed in which a mobile device leaves a heavily jammed/interfered-with frequency or area. It is shown that, by applying reinforcement learning techniques, a mobile device can achieve an optimal communication policy without the need to know the jamming and interference model and the radio channel model in a dynamic game framework. More specifically, a hotbooting deep Q-network based two-dimensional mobile communication scheme is proposed that exploits experiences in similar scenarios to reduce the exploration time at the beginning of the game, and applies deep convolutional neural network and macro-action techniques to accelerate the learning speed in dynamic situations. Several real-world scenarios are simulated to evaluate the proposed method. These simulation results show that our proposed scheme can improve both the signal-to-interference-plus-noise ratio of the signals and the utility of the mobile devices against cooperative jamming compared with benchmark schemes.
\end{abstract}

}

\maketitle

\IEEEdisplaynontitleabstractindextext

%
\IEEEpeerreviewmaketitle

\section{Introduction}\label{sec:introduction}
\IEEEPARstart{B}{y} injecting faked or replayed signals, a jammer aims to interrupt the ongoing communication of mobile devices such as smartphones, mobile sensing robots and mobile servers, and even launch denial of service (DoS) attacks in wireless networks \cite{TMC2,TMC3,berger2016friendly,TMC4,Dams2016Jamming}.
With the pervasion of smart radio devices such as universal software radio peripherals (USRPs) \cite{Rahbari2016Swift}, jammers can cooperatively and flexibly choose their jamming policies to block the mobile devices efficiently \cite{wade2015alone,CBG2015}. 
Jammers can even induce the mobile device to enter a specific communication mode and then launch the jamming attacks accordingly.

Mobile devices usually apply spread spectrum techniques, such as frequency hopping and direct-sequence spread spectrum to address jamming attacks \cite{SurveyofSecurityChallenges,SurveyonSecurityThreats}. However, if most frequency channels in a location are blocked by jammers and/or strongly interfered by electric appliances such as microwaves and other communication radio devices, spread spectrum alone cannot increase the signal-to-interference-plus-noise ratio (SINR) of the received signals.

Therefore, we develop a two-dimensional (2-D) anti-jamming mobile communication system that applies both frequency hopping and user mobility to address both jamming and strong interference. This system has to make a tradeoff between the communication efficiency and the security cost, because a much higher cost is required for the mobile devices to change the geographical location before finishing the communication task than to change the frequency channel, and the overall transmission cost depends on the power allocation scheme of mobile devices. In addition, the mobile device in cognitive radio networks (CRNs) has to avoid interfering with the communication of the primary user (PU).

The anti-jamming communication decisions of a mobile device in a dynamic game can be formulated as a Markov decision process (MDP). Therefore, reinforcement learning (RL) techniques such as Q-learning can be used to achieve an optimal communication policy \cite{1998sutton} in the dynamic game if the jamming and interference model and the radio channel model are not readily available.

The Q-learning based 2-D anti-jamming mobile communication scheme maintains a quality function or Q-function for each state-strategy pair to choose the transmit power and determine whether to leave the location to resist jamming and strong interference via trail-and-error. However, the Q-learning based 2-D mobile communication scheme suffers from the curse of high-dimensionality, i.e., the learning speed is extra slow, if the mobile device has a large number of frequency channels and can observe a large range of the feasible SINR levels.

Therefore, we apply the deep Q-networks (DQN), a deep reinforcement learning technique developed by Google DeepMind \cite{Mnih2015human} to accelerate the learning speed of the 2-D anti-jamming mobile communication system. More specifically, a deep convolutional neural network (CNN) is used for suppressing the state space observed by the mobile device in the dynamic anti-jamming communication game and thus improves the communication performance against jamming and strong interference.

This fast DQN-based communication system applies the macro-action technique in \cite{durugkar2016deep} to further improve the learning speed, which combines the power allocation and mobility decisions in a number of time slots as macro-actions and explores their quality values as a whole. This system also designs a hotbooting technique that exploits the anti-jamming communication experiences in similar scenarios to initialize the CNN weights in the DQN and thus saves exploitation at the beginning of the game.

Simulations are performed to evaluate the performance of our proposed scheme against jammers and interference sources in three mobile communication scenarios: (1) The command dissemination of a mobile server to smart devices such as smart TVs against random jammers, sweep jammers and interference sources, (2) The sensing report transmission of a mobile sensing robot to a server via several access points (APs) against jamming and interference, and (3) The sensing report transmission against two mobile jammers that randomly change their locations. Simulation results show that our proposed mobile communication scheme outperforms the Q-learning based scheme with a faster learning speed, a higher SINR of the signals and a higher utility.

The main contributions of this paper are summarized as follows:
{\begin{itemize}
\item We provide a 2-D anti-jamming mobile communication scheme to resist cooperative jamming and strong interference by applying both frequency hopping and user mobility without being aware of the jamming and interference model and the radio channel model.

\item We apply the communication scheme in the command dissemination of a mobile server to radio devices and the sensing report transmission of a mobile sensing robot against both jamming and interference.

\item We propose a fast DQN-based 2-D mobile communication algorithm that applies DQN, macro-actions and hotbooting techniques to accelerate the learning speed of the Q-learning based communication. Simulation results show that our proposed scheme can increase both the SINR and the utility of the mobile device.
\end{itemize}}

The rest of this paper is organized as follows. We review related work in Section \ref{sec:related} and present the system model in Section \ref{sec:model}. We present the DQN-based 2-D anti-jamming communication system in Section \ref{sec:DQN}, and propose a fast DQN-based communication system in Section \ref{sec:fast}. We provide simulation results in Section \ref{sec:evaluation} and conclude this work in Section \ref{sec:conclusion}.

%
%
%
%

\section{Related Work}\label{sec:related}
Game theory has been applied to study the power allocation of the anti-jamming in wireless communication.
For instance, the Colonel Blotto anti-jamming game presented in \cite{ Wu2012Anti } provides a power allocation strategy to improve the worst-case performance against jamming in cognitive radio networks. The power control Stackelberg game as presented in \cite{Xiao2015Power} formulates the interactions among a source node, a relay node and a jammer that choose their transmit power in sequence without interfering with primary users. The transmission Stackelberg game developed in \cite{reactive2015} helps build a power allocation strategy to maximize the SINR of signals in wireless networks. The prospect-theory based dynamic game in \cite{Xiao2015User} investigates the impact of the subjective decision making process of a smart jammer in cognitive networks under uncertainties. The stochastic game formulated in \cite{SIP2016Optimal} investigates the power allocation of a user against a jammer under uncertain channel power gains.

Game theory has been used for providing insights on the frequency channel selection against jamming. For instance, the stochastic channel access game investigated in \cite{Wang2011An} helps a user to choose the control channel and the data channel to maximize the throughput against jamming. The Bayesian communication game in \cite{wade2016} studies the channel selection against smart jammers with unknown types of intelligence. The zero-sum game as proposed in \cite{Hanawal2015Joint} investigates the frequency hopping and the transmission rate control to improve the average throughput against jamming.

Reinforcement learning techniques enable an agent to achieve an optimal policy via trials in Markov decision process. 
The Q-learning based power control strategy developed in \cite{Xiao2015Power} makes a tradeoff between the defense cost and the communication efficiency without being aware of the jamming model. The Q-learning based channel allocation scheme as proposed in \cite{Gwon2013Competing} can achieve an optimal channel access strategy for a radio transmitter with multiple channels in the dynamic game.
The synchronous channel allocation in \cite{Slimeni2015Jamming} applies Q-learning to proactively avoid using the blocked channels in cognitive radio networks. The multi-agent reinforcement learning (MARL) based channel allocation as proposed in \cite{Lo2012Multiagent} enhances the transmission and sensing capabilities for cognitive radio users. The MARL-based power control strategy as developed in \cite{He2015Faster} accelerates the learning speed of the energy harvesting communication system against intelligent adversaries.

The 2-D anti-jamming mobile communication system proposed in \cite{two2017} uses both frequency and spatial diverting to improve the communication performance against jamming and applies DQN to derive an optimal policy without knowing the jamming and interference model and the radio channel model.

In this work, we present a fast DQN-based power and mobile control scheme that applies the hotbooting and macro-actions techniques to accelerate the learning speed and thus improve the jamming resistance of the communication scheme as proposed in \cite{two2017} for the mobile communication system with a large number of channels. We investigate the applications of this scheme in the sensing report transmission of a mobile sensing robot and the command dissemination of a mobile server to the smart devices against jamming and interference. We evaluate the performance of our proposed schemes against both static and mobile jammers in the sensing report transmission.


\section{System Model}\label{sec:model}
\subsection{Network Model}
\begin{figure}[!t]
\begin{center}
\includegraphics[height=2.5 in]{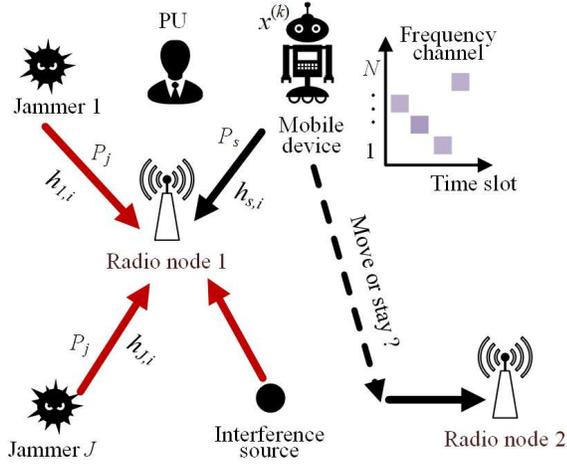}\\
\caption{Network model of the 2-D anti-jamming communication of a mobile device with $N$ frequency channels, against $J$ jammers and interference sources.}\label{System}
\end{center}
\end{figure}

A mobile device such as a smartphone, a mobile server or a mobile sensing robot is aims to transmit messages with high SINR in a frequency hopping based system to serving radio nodes such as an AP or smart devices against $J$ jammers and interference sources.

As shown in Fig. \ref{System}, the mobile device chooses the frequency channel denoted by $f^{(k)}$, the transmit power denoted by $P_s^{(k)}$ and whether to move its location denoted by $\phi^{(k)}$ at time $k$. The channel $f^{(k)}$ is selected according to the frequency patterns shared with the serving radio node. The feasible transmit power $P_s^{(k)}$ is quantized into $L+1$ levels, i.e., $P_s^{(k)}\in \left[Pl/L\right]_{0\leq l \leq L}$, where $P$ is the maximum transmit power. The mobile device stays in the location if $\phi^{(k)}=0$ and moves to another location to connect to another radio node, if $\phi^{(k)}=1$.

Both the mobile device and the radio node choose the frequency pattern $C_{\psi}=\left[c_{\psi}^{(i)}\right]_{1 \leq i \leq \kappa}$, where $c_{\psi}^{(i)} \in \left[1,\cdots,N\right]$ is the index of the frequency channel at time $i$, and $\kappa$ is the number of time slots in the frequency pattern, with $1 \leq \psi \leq \vartheta$, where $\vartheta$ is the number of feasible frequency patterns. The mobile device sends a message to the radio node at time $k$ over the frequency channel $c_{\psi}^{k \bmod \kappa\ +1}$.

Upon receiving the message, the serving radio node evaluates the bit error rate (BER) of the message and estimates the SINR of the signals and chooses the frequency pattern index $\psi^{(k)}$ which are sent to the mobile device on the feedback channel.

The mobile device has to avoid interfering with the communication of PU if in a cognitive radio network. The absence of the PU is denoted by $\lambda^{(k)}$, which equals $0$ if the mobile device detects a PU accessing channel $f^{(k)}$ in the location and $1$ otherwise. The mobile device applies a spectrum sensing technique, such as the energy detection as presented in \cite{2009Cognitive} to detect the PU presence and thus obtains $\lambda^{(k)}$. Let the channel vector $\mathbf{h}_s^{(k)}=\left[h_{s,i}^{(k)}\right]_{1\leq i \leq N}$ denote the channel power gains of the $N$ channels from the mobile device to the serving radio node, $C_{h}$ be the cost of frequency hopping to the  mobile device, $C_p$ be the unit transmission cost and $C_{m}$ be the extra cost of user mobility.

\subsection{Jamming Model}
A jammer sends replayed or faked signals with power $P_J$ on the selected jamming channels to interrupt the ongoing communication of the mobile device. If failing to do that, the jammer also aims to reduce the SINR of the signals received by the radio node with less jamming power. We will consider three types of jamming attacks similar to \cite{2009Efficient} and \cite{Wilhelm2011Short}:
{\begin{itemize}
\item A random jammer randomly selects a jamming channel in each time slot, using the same jamming channel with probability $1-\epsilon$ and a new channel with probability $\epsilon$.
\item A sweep jammer blocks $N_J$ neighboring channels in each time slot from the $N$ channels in sequence and each channel is jammed with power $P_J/N_J$.
\item A reactive jammer as the most harmful chooses the jamming policy based on the the ongoing communication \cite{Wilhelm2011Short}. In this work, we assume that a reactive jammer monitors $N_r$ channels with the energy detection and attacks the active channel.
\end{itemize}}

The jamming channel chosen by jammer $j$ at time $k$ denoted by $y_{j}^{(k)}$ $\in \left[1,\cdots,N\right]$. For simplicity, we define the action set of the $J$ jammers at different locations in the area $\mathbf{y}^{(k)}=\left[y_{j}^{(k)}\right]_{1\leq j \leq J}$. By applying smart and programable radio devices, these jammers sometimes can block all the radio channels if the serving node is close enough to the jammer.

The status of the interference source at time $k$ is denoted by $\eta^{(k)}$, which equals 1 if it interferes the ongoing message transmission of the mobile device with power $P_{f}$ and 0 otherwise. The receiver noise power is denoted by $\sigma$. The channel power gains from the $J$ jammers to the serving radio node on the $N$ channels are denoted by $\mathbf{h}_j^{(k)} = \left[h_{j,i}^{(k)}\right]_{1\leq j \leq J, 1\leq i \leq N}$.

For simplicity, we assume that the static jammer cannot block the mobile device from the new location with another serving radio node and the jammer cannot obtain the frequency patterns nor block the feedback channel. On the other hand, this scheme is applicable to the case with a smart jammer that can attack the feedback channel. For instance, if the mobile device fails to receive the feedback signal, it can set the SINR to be zero.

\subsection{Game Model}

\begin{table}[!b]
\renewcommand{\arraystretch}{1.1}
  \caption{SUMMARY OF SYMBOLS AND NOTATIONS}
\newcommand{\tabincell}[2]{\begin{tabular}{@{}#1@{}}#2\end{tabular}}
  \centering
  \begin{tabular}{|c l|}\hline
\textbf{Notation} & \tabincell{l}{\textbf{Description}}\\ \hline
$N$ & \tabincell{l}{Number of frequency channels } \\ \hline
$J$ & \tabincell{l}{Number of jammers} \\ \hline
$P_{s}$ & \tabincell{l}{Transmit power of the mobile device}\\ \hline
$P_{J/f}$ & \tabincell{l}{Jamming / interference power}\\ \hline
$f^{(k)}$ & \tabincell{l}{The chosen channel at time $k$}\\ \hline
$\psi^{(k)}$ & \tabincell{l}{The chosen frequency pattern index}\\ \hline
$\phi^{(k)}$ & \tabincell{l}{User mobility indicator}\\ \hline
$P$ & \tabincell{l}{The maximum transmit power}\\ \hline
$\vartheta$ & \tabincell{l}{Number of frequency patterns}\\ \hline
$\kappa$ & \tabincell{l}{Length of a frequency pattern}\\ \hline
$N_J$ & \tabincell{l}{Number of channels for sweep jammer}\\ \hline
$N_r$ & \tabincell{l}{Number of channels for reactive jammer}\\ \hline
$\mathbf{h}_{s}^{(k)}$ & \tabincell{l}{Channel power gains of the mobile device} \\ \hline
$\mathbf{h}_{j}^{(k)}$ & \tabincell{l}{Channel power gains of jammer} \\ \hline
$\lambda^{(k)}$ & \tabincell{l}{Absence of PU at time $k$} \\ \hline
$\eta^{(k)}$& \tabincell{l}{Status of the interference source }\\ \hline
$\sigma$ & \tabincell{l}{Receiver noise power} \\ \hline
$C_{m}$ & \tabincell{l}{Cost of user mobility} \\ \hline
$C_{h}$ & \tabincell{l}{Cost of frequency hopping} \\ \hline
$C_{p}$ & \tabincell{l}{Unit transmission cost} \\ \hline
$u^{(k)}$ & \tabincell{l}{Utility of the mobile device} \\ \hline
$\textbf{s}^{(k)}$ & \tabincell{l}{System state} \\ \hline
$\xi$ & \tabincell{l}{Number of the SINR quantization levels} \\ \hline
$\gamma$ & \tabincell{l}{Discount factor in the learning algorithm} \\ \hline
$W$ & \tabincell{l}{Size of the state-action pairs in the CNN} \\ \hline
$\bm{\varphi}^{(k)}$ & \tabincell{l}{State sequence at time $k$} \\ \hline
$\boldsymbol{\theta}^{(k)}$ & \tabincell{l}{CNN weights at time $k$} \\ \hline
$\textbf{e}^{(k)}$ & \tabincell{l}{Experience at time $k$ } \\ \hline
$\alpha$ & \tabincell{l}{Learning rate} \\ \hline
$B$ & \tabincell{l}{Size of the CNN minibatch} \\ \hline
$\mathcal{M}$ & \tabincell{l}{Macro-actions set} \\ \hline
$\Phi$ & \tabincell{l}{Number of the macro-actions} \\ \hline
$\zeta$ & \tabincell{l}{Length of a macro-action} \\ \hline
$p$ & \tabincell{l}{Jammer mobility probability} \\ \hline
\end{tabular}
\end{table}

The repeated interactions between the mobile device and the jammer are formulated as a zero-sum game. The mobile device chooses a communication strategy denoted by $\bm{x}^{(k)}=\left[P_s^{(k)},\phi^{(k)}\right]\in \mathcal{X}$, where $\mathcal{X}$ is the all possible strategies set, and selects a frequency channel $f^{(k)}$ to send signals and determines whether to change the location. Each jammer chooses the jamming channel. The utility of the mobile device at time $k$ in the game, denoted by $u^{(k)}$, depends on the SINR of the received signals and the communication costs which consist of the user mobility cost, the transmission cost and the channel hopping cost.

Let $\mathcal{F}(\varsigma)$ be an indicator function that equals $1$ if $\varsigma$ is true, and $0$ otherwise. The utility of the mobile device is given by
\begin{align}\label{Epus}
u^{(k)}&\left(\bm{x},\mathbf{y}\right)=\frac{P_s^{(k)}h_{s,f}^{(k)}\lambda^{(k)}}{\sigma+P_{f}\eta^{(k)}+\sum_{j=1}^{J}P_Jh_{j,y_j}^{(k)}\mathcal{F}\left(f^{(k)}==y_j^{(k)}\right)} \nonumber \\ &-C_pP_s^{(k)}-C_{m}\phi^{(k)}-C_{h}\mathcal{F}\left(f^{(k)}\neq f^{(k-1)}\right).
\end{align}
For simplicity, both the $J$ jammers and the interference sources are assumed to be close to the serving radio node, although this work can be extended to other cases. The proposed anti-jamming communication systems in the next sections also apply in the case with other utility formulation. As the next system state observed by the mobile device is independent of the previous system state and action, making this anti-jamming game as a Markov decision process. For ease of reference, important notations are summarized in Table 1.

\section{DQN-based 2-D anti-jamming mobile communication scheme}\label{sec:DQN}
\begin{figure*}[!t]
\begin{center}
\includegraphics[height=3.7 in]{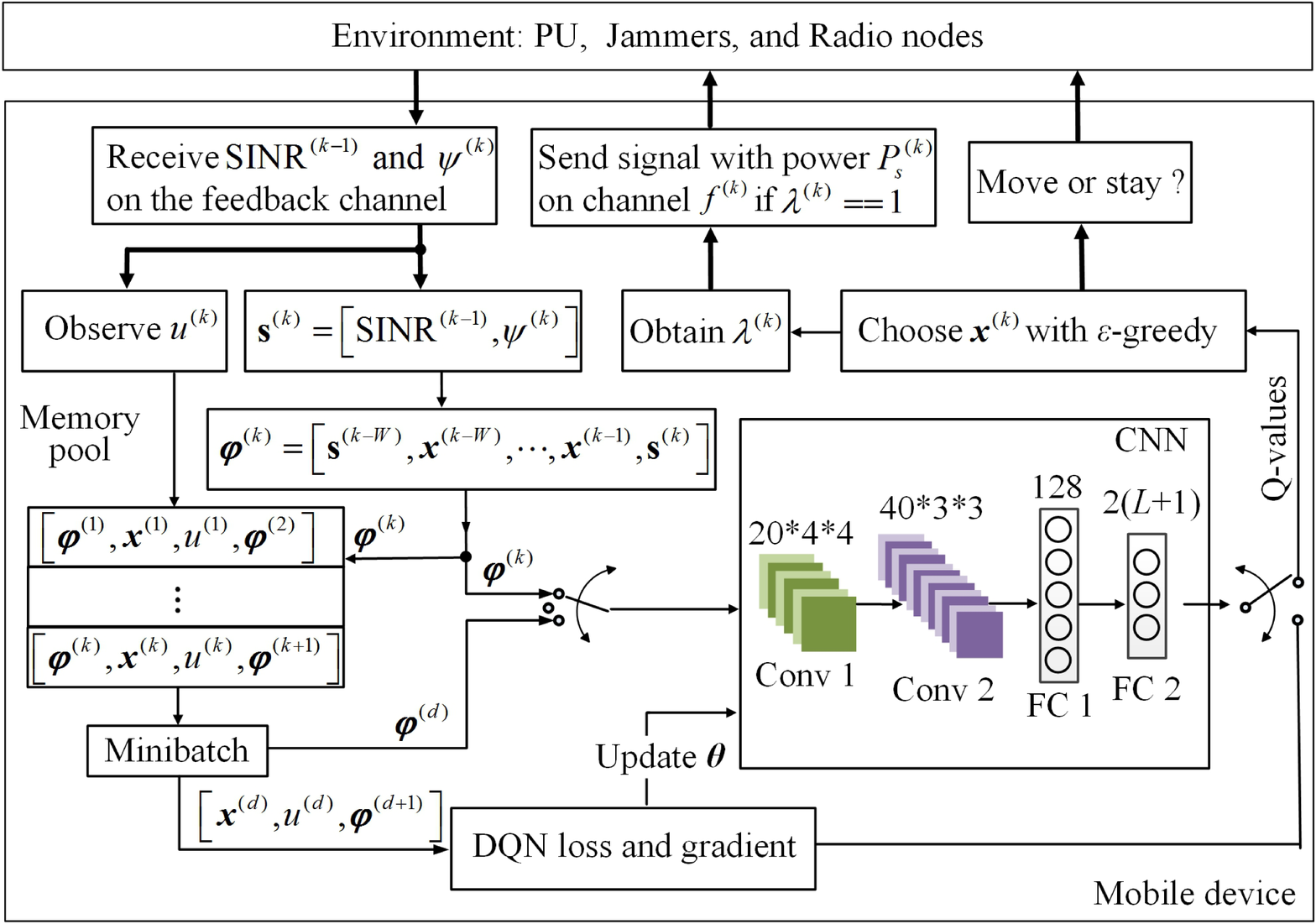}\\
\caption{DQN-based $2$-D anti-jamming mobile communication scheme.}\label{DQN}
\end{center}
\end{figure*}

Without being aware of the jamming and interference model and the radio channel model, a mobile device can apply reinforcement learning techniques such as Q-learning to achieve an optimal communication policy via trial-and-error in the dynamic game.

The learning speed of the Q-learning based 2-D communication algorithm proposed in our previous work in \cite{two2017} suffers from the curse of high-dimensionality, i.e., the required convergence time increases with the dimension of the state space and the feasible communication strategy set, which increases with the number of frequency channels and the power quantization levels used by the mobile device. Therefore, we proposed a 2-D mobile communication scheme based on deep Q-network, a deep reinforcement learning technique that applies deep convolutional neural network to compress the state space observed by the mobile device. The mobile device receives the SINR of the signals and the chosen frequency pattern index on the feedback channel and formulates the system state denoted by $\textbf{s}$ as $\textbf{s}^{(k)}=\left[{\rm SINR}^{(k-1)},\psi^{(k)}\right] \in \textbf{S}$, where $\textbf{S}$ is the feasible state set. The size of the system state set is $|\textbf{S}|=\vartheta\xi$, where $\xi$ is the number of the SINR quantization levels.

As illustrated in Fig. \ref{DQN}, the communication strategy of the mobile device is chosen based on the quality function or Q-function of the current system state, which is the expected discounted long-term reward for each state-strategy pair, and defined as
\begin{align}\label{quality}
Q(\textbf{s},\bm{x})=\mathbb{E}_{\textbf{s}' \in \textbf{S}}\left[u^{(k)}+\gamma \max_{ \bm{x}' \in \mathcal{X}}{Q\Big(\textbf{s}',\bm{x}'\Big)}\bigg|\textbf{s},\bm{x}\right],
\end{align}
where $\textbf{s}'$ is the next state if the mobile device takes strategy $\bm{x}$ at state \textbf{s}, and the discount factor $\gamma$ represents the uncertainty of the mobile device regarding the future reward in the dynamic game against jamming and interference.

\begin{table}[!b]
\small
\renewcommand{\arraystretch}{1.2}
\centering
\caption{CNN parameters in the mobile communication scheme in Algorithm 1}
\begin{tabular}{|c|c|c|c|c|}
\hline
\textbf{Layer} & \textbf{Conv 1} & \textbf{Conv 2} & \textbf{FC 1} & \textbf{FC 2}  \\
\hline
\textbf{Input} & $1*6*6$ & $20*4*4$ & 320 & 128 \\
\hline
\textbf{Filter size} & $3*3$ & $2*2$ & \ /  &  \ / \\
\hline
\textbf{Stride} & $1$ & $1$ & \ / &  \ / \\
\hline
\textbf{No. of filters} & $20$ & $40$ & $128$ & $2(L+1)$  \\
\hline
\textbf{Activation} & ReLU & ReLU & ReLU & \ /  \\
\hline
\textbf{Output}  & $20*4*4$ & $40*3*3$ & $128$ & $2(L+1)$  \\
\hline
\end{tabular}
\end{table}

\begin{algorithm}[!b]
\caption {DQN-based 2-D mobile communication}
Initialize $\bm{\theta}^{(0)}$, $\gamma$, $\mathcal{D}=\emptyset$, ${\rm SINR}^{(0)}$, $\psi^{(1)}$, $W$, and $B$\\
\For{$k=1, 2, \cdots $}
{
$\textbf{s}^{(k)}=\left[{\rm SINR}^{(k-1)}, \psi^{(k)} \right]$\\
\uIf{$k\leq W$}
{
Choose $\bm{x}^{(k)}=\left[P_s^{(k)},\phi^{(k)}\right]$ at random
}
\Else
{
Obtain the CNN outputs as the $Q$-values with the input $\bm{\varphi}^{(k)}$\\
Choose $\bm{x}^{(k)}$ via (\ref{epsilon-greedy})\\
}
\uIf {$\phi^{(k)}$ == $1$}
{
Change the location and connect to a new radio node\\
}
Observe the absence of PU and set $\lambda^{(k)}$\\
\uIf {$\lambda^{(k)}$ == $0$}
{
Keep silence
}
\Else
{
$f^{(k)}\leftarrow c_\psi^{k \bmod \kappa +1}$\\
Send signals on channel $f^{(k)}$ with power $P_s^{(k)}$
}
Receive the ${\rm SINR}^{(k)}$ and $\psi^{(k+1)}$ on the feedback channel\\
Obtain $u^{(k)}$ and $\textbf{s}^{(k+1)}=\left[{\rm SINR}^{(k)}, \psi^{(k+1)} \right]$\\
Observe $\bm{\varphi}^{(k+1)}=\big\{\textbf{s}^{(k-W+1)},\bm{x}^{(k-W+1)},\cdots,\bm{x}^{(k)},$
$ \textbf{s}^{(k+1)}\big\}$\\
$\mathcal{D} \leftarrow \left\{\bm{\varphi}^{(k)},\bm{x}^{(k)},u^{(k)},\bm{\varphi}^{(k+1)}\right\} \cup \mathcal{D}$\\
\For {$d=1,2,\cdots,B$}
{
Select $\left\{\bm{\varphi}^{(d)},\bm{x}^{(d)},u^{(d)},\bm{\varphi}^{(d+1)}\right\}$ $\in$ $\mathcal{D}$ at random\\
Calculate $R$ via (\ref{Rd})\\
}
Update $\bm{\theta}^{(k)}$ via (\ref{gradient})\\
}
\end{algorithm}

The deep convolutional neural network is used in the DQN-based 2-D mobile communication scheme as a nonlinear function approximator to estimate the value of the Q-function in (\ref{quality}) for each strategy, as the state set size $| \textbf{S} |$ is too large for a Q-learning based scheme to quickly achieve an optimal policy. The state sequence at time $k$ denoted by $\bm{\varphi}^{(k)}$ consists of the current system state and the previous $W$ system state-strategy pairs, i.e., $\bm{\varphi}^{(k)}=\left[\textbf{s}^{(k-W)},\bm{x}^{(k-W)},\cdots,\bm{x}^{(k-1)},\textbf{s}^{(k)}\right]$. As shown in Fig. \ref{DQN}, the state sequence $\bm{\varphi}^{(k)}$ is reshaped into a $6\times6$ matrix and taken as the input to the CNN.

As shown in Fig. \ref{DQN}, the CNN in the DQN-based 2-D mobile communication scheme consists of two convolutional (Conv) layers and two fully connected (FC) layers. The first Conv layer includes $20$ filters, each with size $3 * 3$ and stride $1$ as summarized in Table 2. The second Conv layer has $40$ filters, each with size $2 * 2$ and stride $1$. Both layers use the rectified linear units (ReLU) as the activation function. The first FC layer involves $128$ rectified linear units, and the second FC layer has $2(L+1)$ outputs for each feasible strategy. The filter weights of the four layers in the CNN at time $k$ are denoted by $\boldsymbol{\theta}^{(k)}$, which are updated at each time slot based on the experience replay. The output of the CNN is used for estimating the values of the Q-function for the $2(L+1)$ actions, $Q\left(\bm{\varphi}^{(k)}, \bm{x}| \boldsymbol{\theta}^{(k)}\right)$,$\forall \bm{x} \in \mathcal{X}$.

The DQN-based 2-D mobile communication scheme stores each experience observed by the mobile device denoted by $\textbf{e}^{(k)}=\left\{\bm{\varphi}^{(k)},\bm{x}^{(k)},u^{(k)},\bm{\varphi}^{(k+1)}\right\}$, and maintains a memory pool given by $\mathcal{D}=\left\{\textbf{e}^{(1)},\cdots,\textbf{e}^{(k)}\right\}$. By applying the experience replay, the mobile device chooses an experience $\textbf{e}^{(d)}$ from the memory pool $\mathcal{D}$ at random, with $1 \leq d \leq k$ to update $\boldsymbol{\theta}^{(k)}$. According to a stochastic gradient descent (SGD) algorithm, the mean-squared error of the target optimal Q-function value is minimized with minibatch updates. More specifically, the loss function is chosen by \cite{Mnih2015human} as
\begin{align}\label{theta}
L\Big(\boldsymbol{\theta}^{(k)}\Big)=\mathbb{E}_{\bm{\varphi},\bm{x},u,\bm{\varphi'}}\Bigg[\bigg( R-Q\Big(\bm{\varphi},\bm{x};\boldsymbol{\theta}^{(k)} \Big) \bigg)^2\Bigg],
\end{align}
where the target optimal Q-function $R$ is given by
\begin{align}\label{Rd}
R = u^{(k)}+\gamma \max_{\bm{x}' \in \mathcal{X}}Q\Big(\bm{\varphi'},\bm{x}';\boldsymbol{\theta}^{(k-1)}\Big),
\end{align}
and $\bm{\varphi'}$ is the next state sequence.

The gradient of the loss function with respect to the weights $\boldsymbol{\theta}^{(k)}$ is given by
\begin{align}\label{gradient}
\nabla_{\boldsymbol{\theta}^{(k)}}&L\Big(\boldsymbol{\theta}^{(k)}\Big)=\mathbb{E}_{\bm{\varphi},\bm{x},u,\bm{\varphi'}}\bigg[ R\nabla_{\boldsymbol{\theta}^{(k)}}Q\Big(\bm{\varphi},\bm{x};\boldsymbol{\theta}^{(k)} \Big) \bigg] \nonumber \\
&-\mathbb{E}_{\bm{\varphi},\bm{x}}\bigg[ Q\Big(\bm{\varphi},\bm{x};\boldsymbol{\theta}^{(k)} \Big)\nabla_{\boldsymbol{\theta}^{(k)}}Q\Big(\bm{\varphi},\bm{x};\boldsymbol{\theta}^{(k)}\Big)\bigg].
\end{align}
This process repeats $B$ times and $\boldsymbol{\theta}^{(k)}$ is then updated according to these randomly selected experiences.

The $\epsilon$-greedy algorithm is applied to choose communication strategy to avoid staying in the local maximum at the beginning of the game. More specifically, the optimal strategy with the highest Q-function value is chosen with a high probability $1 - \epsilon$, and other feasible strategies are chosen with small probability, i.e.,
\begin{align}\label{epsilon-greedy}
\mathrm{Pr}\Big(\bm{x}^{(k)}=\dot{\bm{x}}\Big)=
\begin{cases}
1-\epsilon,  &\dot{\bm{x}}=\arg \max \limits_{\bm{x}' \in \mathcal{X}}Q\Big(\bm{\varphi}^{(k)},\bm{x}'\Big)\\
\frac{\epsilon}{2L+1},  & \text{o.w.}
\end{cases}
\end{align}


As summarized in Algorithm 1, the mobile device observes the SINR of the signals from the serving radio node at time $k$ to update the system state, and receives utility $u^{(k)}$. According to the next state sequence $\bm{\varphi}^{(k+1)}$, the new experience $\left\{\bm{\varphi}^{(k)},\bm{x}^{(k)},u^{(k)},\bm{\varphi}^{(k+1)}\right\}$ is stored in the memory pool $\mathcal{D}$.

\begin{algorithm}[!b]
\caption {Hotbooting process}
Initialize $I$, $K$, $W$, $B$, $\mathbb{E}=\emptyset$, and $\bm{\theta}_{*}^{(0)}$\\
\For{$i=1,2,\cdots,I$}
{
Initialize a emulational environment\\
Initialize ${\rm SINR}^{(0)}$ and $\psi^{(1)}$\\
\For{$k=1,2,\cdots,K$}
{
Perform Steps $3-21$ in Algorithm 1 to choose $\bm{x}^{(k)}$, obtain $u^{(k)}$ and observe $\bm{\varphi}^{(k+1)}$ \\
$\mathbb{E} \leftarrow \mathbb{E} \cup \{\bm{\varphi}^{(k)}, \bm{x}^{(k)}, u^{(k)}, \bm{\varphi}^{(k+1)}\}$\\
\For{$d=1,2,\cdots,B$}
{
Select $\left\{\bm{\varphi}^{(d)},\bm{x}^{(d)},u^{(d)},\bm{\varphi}^{(d+1)}\right\}$ $\in$ $\mathbb{E}$ at random\\
Calculate $R$ via (\ref{Rd})
}
Update $\bm{\theta}_{*}^{(k)}$ via (\ref{gradient})
}
}
$\mathbf{Output}$ $\bm{\theta}_{*}$
\end{algorithm}

\begin{algorithm}[!b]
\caption {Fast DQN-based 2-D mobile communication}
Call Algorithm 2\\
$\bm{\theta}^{(0)} \leftarrow \bm{\theta}_*$\\
Initialize $\gamma$, $T$, $\mathcal{D}=\emptyset$, $\zeta$, $\Phi$, $\mathcal{M}=\emptyset$, ${\rm SINR}^{(0)}$, $\psi^{(1)}$ and $\mathbf{\bar{u}}=\mathbf{0}$ \\
\For{$t=1,\cdots,T$}
{
Choose $\bm{x}^{(t)}=\left[P_s^{(t)},\phi^{(t)}\right]$ at random\\
Perform Steps $10-20$ in Algorithm 1 to obtain $u^{(t)}$\\
$\bar{u}\left(\bm{x}^{(t)}\right)\leftarrow \max \left\{u^{(t)}, \bar{u}\left(\bm{x}^{(t)}\right) \right\}$\\
}
$A=\left\{\bm{x}_{1}, \bm{x}_{2},\cdots, \bm{x}_{2L+2} \right\}$, $\forall \bar{u}\left(\bm{x}_i\right) > \bar{u}\left(\bm{x}_j\right)$, $i < j$\\
Store $m_i=\left\{ \bm{x}_{i}^{(1)},\cdots, \bm{x}_{i}^{(\zeta)} \right\}, \forall 1 \leq i \leq \Phi $ in $\mathcal{M}$\\
\For{$k=1, 2, \cdots $}
{
Perform Steps 3-12 in Algorithm 1\\
\uIf {$\lambda^{(k)}$ == $0$}
{
Keep silence
}
\Else
{\uIf{$\bm{x}^{(k)}$ $\in \mathcal{M}$}
{
Follow the macro-action $\bm{x}^{(k)}$ in the next $\zeta$ time slots\\
Obtain $\left[u^{(v)}\right]_{k\leq v \leq k+\zeta-1}$ and observe a series of states $\left[\textbf{s}^{(l)}\right]_{k+1\leq l \leq k+\zeta}$\\
Obtain a series of the CNN inputs $\left[ \bm{\varphi}^{(i)}\right]_{k+1\leq i \leq k+\zeta}$\\
Calculate $U^{(k)}$ via (\ref{U})\\
$\mathcal{D} \leftarrow \left\{\bm{\varphi}^{(k)},\bm{x}^{(k)},U^{(k)},\bm{\varphi}^{(k+\zeta)}\right\} \cup \mathcal{D}$\\
$k \leftarrow k+\zeta$\\
}
\Else
{
Perform Steps 16-22 in Algorithm 1 to obtain $\mathcal{D} \leftarrow \left\{\bm{\varphi}^{(k)},x^{(k)},u^{(k)},\bm{\varphi}^{(k+1)}\right\} \cup \mathcal{D}$\\
}
}

\For{$d=1,2,\cdots,B$}
{
Select $\left(\bm{\varphi}^{(d)},\bm{x}^{(d)},u^{(d)},\bm{\varphi}^{(d+1)}\right) \in \mathcal{D}$ at random\\
\uIf{$\bm{x}^{(d)}$ $\in \mathcal{M}$}
{
Calculate $R$ via (\ref{Rd_M})\\
}
\Else
{
Calculate $R$ via (\ref{Rd})\\
}
}
Update $\bm{\theta}^{(k)}$ via (\ref{gradient})
}
\end{algorithm}

\section{Fast DQN-based 2-D Anti-jamming mobile communication scheme}\label{sec:fast}
The DQN-based 2-D mobile communication scheme as proposed in the last section that estimates the Q-function based on the CNN can apply the anti-jamming communication history to accelerate the learning speed at the beginning of the dynamic game. Therefore, a hotbooting process initializes the CNN parameters $\bm{\theta}^{(0)}$ with the communication experiences instead of the random initial values in Algorithm 1.

We propose a fast DQN-based 2-D mobile communication scheme that uses the hotbooting process as shown in Algorithm 2 to exploit the experiences learnt from the similar communication scenarios. In each scenario, the mobile device chooses a communication strategy $\bm{x} = \left[P_s, \phi\right]$, and observes the resulting SINR of the signals and the utility. Each emulational experience
$\{ \bm{\varphi}^{(k)}, \bm{x}^{(k)}, u^{(k)}, \bm{\varphi}^{(k+1)} \}$ is stored in the database $\mathbb{E}$, which presents how the mobile device chooses the communication strategy $\bm{x}^{(k)}$  at state $\bm{\varphi}^{(k)}$ via (\ref{epsilon-greedy}) to receive utility $u^{(k)}$ and reach a new state $\bm{\varphi}^{(k+1)}$. The mobile device samples the emulational experiences from the database $\mathbb{E}$ to update the CNN parameters $\bm{\theta}_{*}$ according to the SGD algorithm via (\ref{gradient}).

The fast DQN-based 2-D mobile communication scheme also applies the temporal abstraction to accelerate the learning speed for the case with a large action space. The mobile device takes hierarchical multi-step actions as macro-actions or macros at different timescales. The macros are deterministic sequences of the power allocation and mobility decisions, i.e., a macro-action $m=\left[ \bm{x}^{1},\cdots,\bm{x}^{\zeta} \right] \in \mathcal{M}$, where $\mathcal{M}$ is the set of all macros and $\zeta$ is the length of a macro-action.

The mobile device transmits a message with a randomly chosen communication strategy $\bm{x}$ and evaluates the SINR and the utility. All the communication strategy experiences are sorted according to the utility. The top $\Phi$ communication strategies are chosen to construct the macros. Each macro-action $m$ consists of the same strategy $\zeta$ time slots in sequence. After applying macros, the mobile device updates the number of the CNN outputs to $2(L+1)+\Phi$.

\begin{algorithm}[!b]
\caption {Q-learning based 2-D mobile communication}
Initialize $\gamma$, $\alpha$, $P_s$, ${\rm SINR}^{(0)}$, $\psi^{(1)}$, $\mathbf{Q}=\mathbf{0}$, and $\mathbf{V}=\mathbf{0}$\\
\For{$k = 1, 2, \cdots$}
{
$\textbf{s}^{(k)}=\left[{\rm SINR}^{(k-1)}, \psi^{(k)} \right]$\\
Choose $\bm{x}^{(k)}=\left[P_s^{(k)},\phi^{(k)}\right]$ via (\ref{epsilon-greedy})\\
\uIf {$\phi^{(k)}$ == $1$}
{
Change the location and connect to a new radio node\\
}
Observe the absence of PU and set $\lambda^{(k)}$\\
\uIf {$\lambda^{(k)}$ == $0$}
{
Keep silence
}
\Else
{
$f^{(k)}\leftarrow c_\psi^{k \bmod \kappa +1}$\\
Send signals on channel $f^{(k)}$ with power $P_s^{(k)}$
}

Receive the ${\rm SINR}^{(k)}$ and $\psi^{(k+1)}$ on the feedback channel\\

Obtain $u^{(k)}$ and $\textbf{s}^{(k+1)}=\left[{\rm SINR}^{(k)}, \psi^{(k+1)}\right]$\\
Update $Q\Big(\textbf{s}^{(k)},\bm{x}^{(k)}\Big)$ via (\ref{Qs})\\
Update $V\Big(\textbf{s}^{\left(k\right)}\Big)$ via (\ref{Vs})
}
\end{algorithm}

Once a macro-action is chosen, the mobile device will transmit the message by following the communication strategy sequence which is predefined by the macro-action, observe a series of states $\left[\textbf{s}^{(l)}\right]_{k+1\leq l \leq k+\zeta}$ and evaluate the utility sequence $\left[u^{(v)}\right]_{k\leq v \leq k+\zeta-1}$. The optimal target Q-function in (\ref{Rd}) in the fast DQN has to include the macros and is updated according to the cumulative discounted reward \cite{Mcgovern1998Macro}.
More specifically, during a multi-step transition from state $\textbf{s}^{(k)}$ to state $\textbf{s}^{(k+\zeta)}$ with macro-action $m$, the approximate optimal target Q-function with macros is updated by
\begin{align}\label{Rd_M}
R = U^{(k)}+\gamma^{\zeta} \max_{\bm{x}'\in \mathcal{X}}Q\Big(\bm{\varphi}^{(k+\zeta)},\bm{x}';\bm{\theta}^{(k-1)}\Big),
\end{align}
where $U^{(k)}$ is the cumulative discounted reward defined as
\begin{align}\label{U}
U^{(k)}=\sum_{i=0}^{\zeta-1}\gamma^{i}u^{k+i}.
\end{align}

\begin{figure}[!t]\setlength{\abovecaptionskip}{0.cm}\setlength{\belowcaptionskip}{-0.5cm}
\begin{center}
\includegraphics[height=2.4 in]{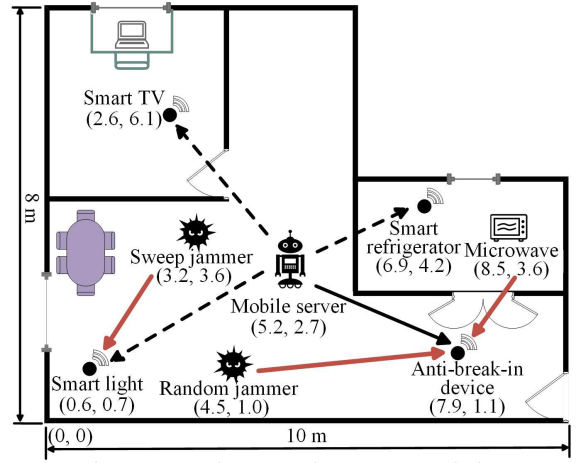}\\
\caption{Simulation topology in the command dissemination of a mobile server against a random jammer, a sweep jammer and an interference source.}\label{EX1}
\end{center}
\end{figure}

The fast DQN-based 2-D anti-jamming mobile communication summarized in Algorithm 3 applies the hotbooting process and the macro-actions technique to improve both the learning speed and the utility in the dynamic game. 

\section{Application and Performance evaluation}\label{sec:evaluation}
We propose a Q-learning based $2$-D anti-jamming mobile communication scheme as summarized in Algorithm $4$, which updates the Q-function according to the iterative Bellman equation as follows:
\begin{align}\label{Qs}
&Q(\textbf{s},\bm{x})\leftarrow \alpha\big(u+\gamma V\left(\textbf{s}'\right)\big)+(1-\alpha)Q(\textbf{s},\bm{x}) \\
\label{Vs}
&V\left(\textbf{s}\right)\leftarrow \max_{\bm{x}' \in \mathcal{X}}Q\big(\textbf{s},\bm{x}'\big),
\end{align}
where $\alpha$ is the learning rate that represents the weight of the current Q-function. 
\begin{figure}[!t]\setlength{\abovecaptionskip}{0.cm}\setlength{\belowcaptionskip}{-0.5cm}
\begin{center}
\includegraphics[height=2.65 in]{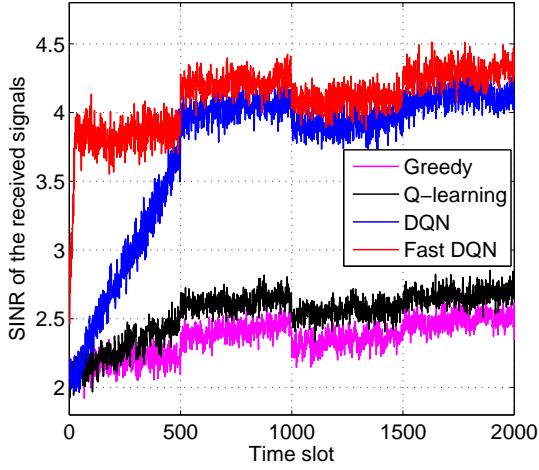}\\
{\footnotesize  (a) SINR of the mobile server signals}\\
\includegraphics[height=2.65 in]{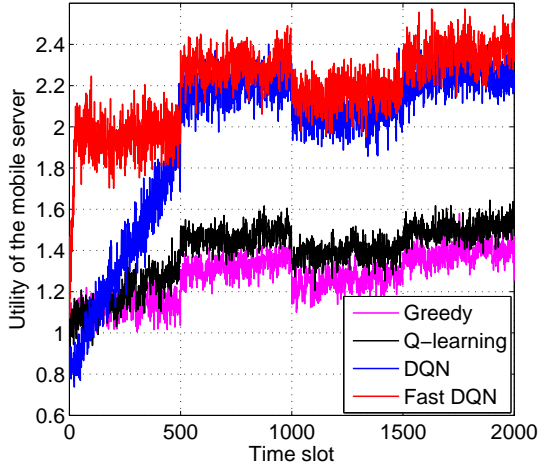}\\
{\footnotesize  (b) Utility of the mobile server}\\
\caption{Performance of the anti-jamming communication scheme in the commands dissemination of a mobile server with 96 frequency channels in a dynamic game against a random jammer, a sweep jammer and an interference source with $C_p=0.2$ in the apartment as shown in Fig. \ref{EX1}.
}\label{SINR_Utility_time_EX1}
\end{center}
\end{figure}

As a benchmark, we propose a greedy-based $2$-D anti-jamming mobile communication scheme, which update the score of each feasible communication strategy according to the average utility it ever achieved.

\subsection{Command dissemination of a mobile server}

The first application of the 2-D mobile communication scheme is the command dissemination of a mobile device in an apartment to 4 smart devices such as an anti-break-in device at the door, a smart TV and a smart refrigerator as shown in Fig. \ref{EX1}. The mobile server chose communication strategy $\bm{x}$ at a given time slot to send command messages to a device according to the settings of the apartment owner against a random jammer, a sweep jammer and an interference source.

The random jammer selects a jamming channel at random, using the same jamming channel with probability 0.9 and a new channel with probability with 0.1. The sweep jammer blocks $N_J=4$ channels simultaneously in each time slot and jams the next $N_J$ channels in the next time slot, with each channel is jammed with power $P_J/N_J$. A microwave in the kitchen sent interference signals during the transmission of the mobile server at a probability 0.05. The channel power gain $\mathbf{h}_s$ changes every 500 time slots and each time slot lasts 10.08 ms.

\begin{figure}[!t]\setlength{\abovecaptionskip}{0.cm}\setlength{\belowcaptionskip}{-0.5cm}
\begin{center}
\includegraphics[height=2.65 in]{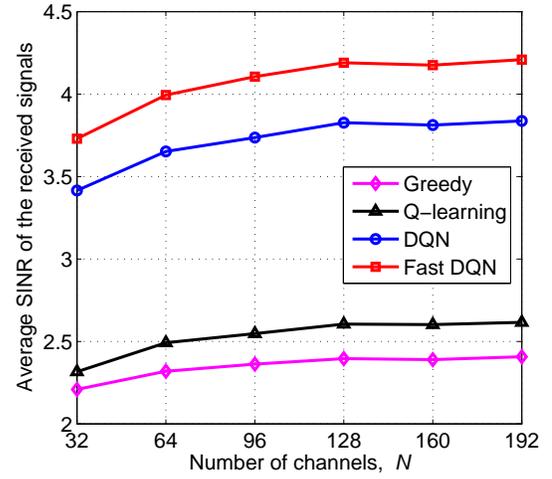}\\
{\footnotesize  (a) Average SINR of the mobile server signals}\\
\includegraphics[height=2.65 in]{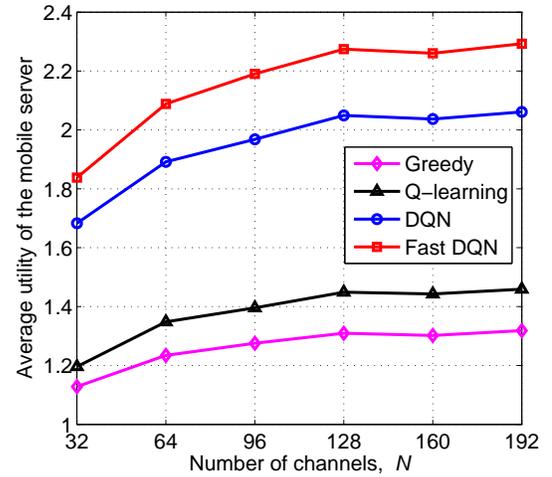}\\
{\footnotesize  (b) Average utility of the mobile server}\\
\caption{Average performance of the anti-jamming communication scheme in the commands dissemination of a mobile server with $N$ frequency channels over 2000 time slots in each dynamic game and 200 scenarios against a random jammer, a sweep jammer and an interference source with $C_p=0.2$ in the apartment as shown in Fig. \ref{EX1}.
}\label{Avg_performance_EX1}
\end{center}
\end{figure}

Simulations have been performed with mobile devices each with Intel i5-6200U CPU, 4GB RAM, and Ubuntu 14.04 64-bits system, and a primary user that randomly chose a channel in each time slot, with $\sigma=1$, $C_{m}=0.8$, $C_p=0.2$, $C_{h}=0.4$, $\mathbf{h}_{s}^{(k)} \in[0,1]$, $\mathbf{h}_{j}^{(k)} \in[0,1]$, $I=200$, $K=200$, $T=300$, $W=11$, $N_r=8$, $N_j=4$, $L=16$, $P=8$, $P_j=8$, $\kappa=30$, $\vartheta=10$, $\Phi=4$ and $\zeta=5$. According to the hyper parameters setting in \cite{Mnih2015human}, we set the minibatch size $B=32$, $\epsilon$ linearly annealed from 0.5 to 0.05 and the learning rate $\alpha$ linearly annealed from 0.7 to 0.5 during the first 300 time slots to accelerate exploration. The discount factor $\gamma$ linearly increased from 0.5 to 0.7 during the first 300 time slots to optimize exploitation and fixed at 0.7 afterwards.

The fast DQN-based scheme outperforms the DQN-based, the Q-learning based and the greedy-based schemes with a higher SINR of the signals and a higher utility due to a faster learning speed as shown in Fig. \ref{SINR_Utility_time_EX1}. For instance, the fast DQN-based scheme increases the SINR of the signals by $31.9$\% compared with the DQN-based scheme, which is $76.2$\% and $84.7$\% higher than that of the Q-learning based and the greedy-based schemes at $300$-th time slot, respectively. Consequently, as shown in Fig. \ref{SINR_Utility_time_EX1}(b), the fast DQN-based scheme improves the utility by $42.4$\%, $80.8$\% and $92.1$\% compared with the DQN-based, the Q-learning based and the greedy-based schemes at that time slot, respectively.

\begin{figure}[!t]\setlength{\abovecaptionskip}{0.cm}\setlength{\belowcaptionskip}{-0.5cm}
\begin{center}
\includegraphics[height=2.65 in]{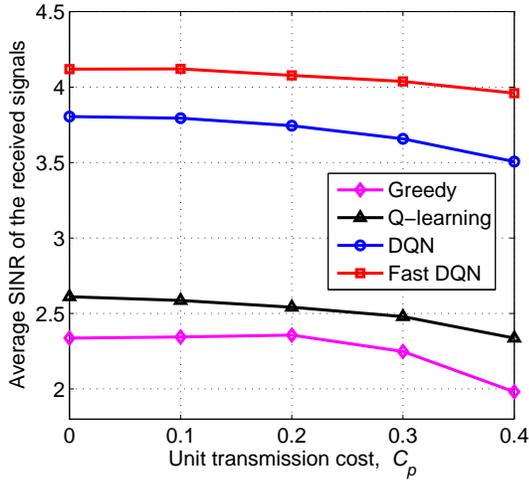}\\
{\footnotesize  (a) Average SINR of the mobile server signals}\\
\includegraphics[height=2.65 in]{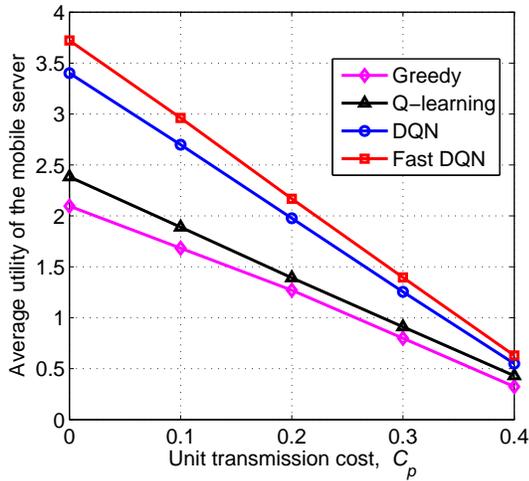}\\
{\footnotesize  (b) Average utility of the mobile server}\\
\caption{Average performance of the anti-jamming communication scheme in the commands dissemination of a mobile server with 96 frequency channels over 2000 time slots in each dynamic game and 200 scenarios against a random jammer, a sweep jammer and an interference source in the apartment as shown in Fig. \ref{EX1}.
}\label{Avg_deffhj_EX1}
\end{center}
\end{figure}
The anti-jamming performance of the proposed scheme improves with the number of channels as shown in Fig. \ref{Avg_performance_EX1}. For example, the average SINR of the signals and the average utility of the mobile server are increased by the DQN-based scheme by $12.1$\% and $21.8$\%, respectively, if the number of channels increases from 32 to 128. In addition, the DQN-based scheme has much better performance than the Q-learning based and the greedy-based schemes and the fast DQN-based scheme can further improve the performance compared with the DQN-based scheme. For instance, the DQN-based scheme achieves $46.7$\% higher SINR and $41.0$\% higher utility compared with the Q-learning based scheme for the system with 96 channels. Furthermore, the fast DQN-based scheme increases the SINR of the signals by $73.8$\% and increases $71.7$\% utility, compared with the greedy-based scheme for the system with 96 channels. On the other hand, the communication efficiency of the RL-based communication scheme has to address the curse of the high-dimensionality under a large number of channels. For instance, the SINR and the utility of all the RL-based schemes no longer improve with $N$ if $N>128$ as shown Fig. \ref{Avg_performance_EX1}.

As shown in Fig. \ref{Avg_deffhj_EX1}, both the SINR of the signals and the utility of the mobile server decrease with the unit transmission cost. For instance, the SINR of the signals and the utility of the mobile server decrease by the DQN-based scheme by $3.9$\% and $63.1$\%, respectively, for the system with $C_p = 0.3$ instead of  $C_p = 0$. In addition, the fast DQN-based strategy always significantly outperforms other three schemes with different $C_p$. For instance, the fast DQN-based scheme increases the SINR of the signals by $75.8$\% compared with the greedy-based scheme, which is $59.3$\% and $8.6$\% higher than that of the Q-learning based and the DQN-based schemes with $C_p = 0.1$, respectively. The fast DQN-based scheme achieves $76.1$\%, $56.8$\% and $9.7$\% higher utility compared with the greedy-based, the Q-learning based and the DQN-based schemes, respectively.

In the simulation, the mobile device takes on average 2ms to update CNN weight parameters and choose the communication strategy. The data size is 100KB and the signal rate is 100Mb/s, the average transmission latency is 8ms, if the feedback time is 0.08ms and the feedback data size is 1KB.

\subsection{Sensing report collection}
\begin{figure}[!t]\setlength{\abovecaptionskip}{0.cm}\setlength{\belowcaptionskip}{-0.5cm}
\begin{center}
\includegraphics[height=2.4 in]{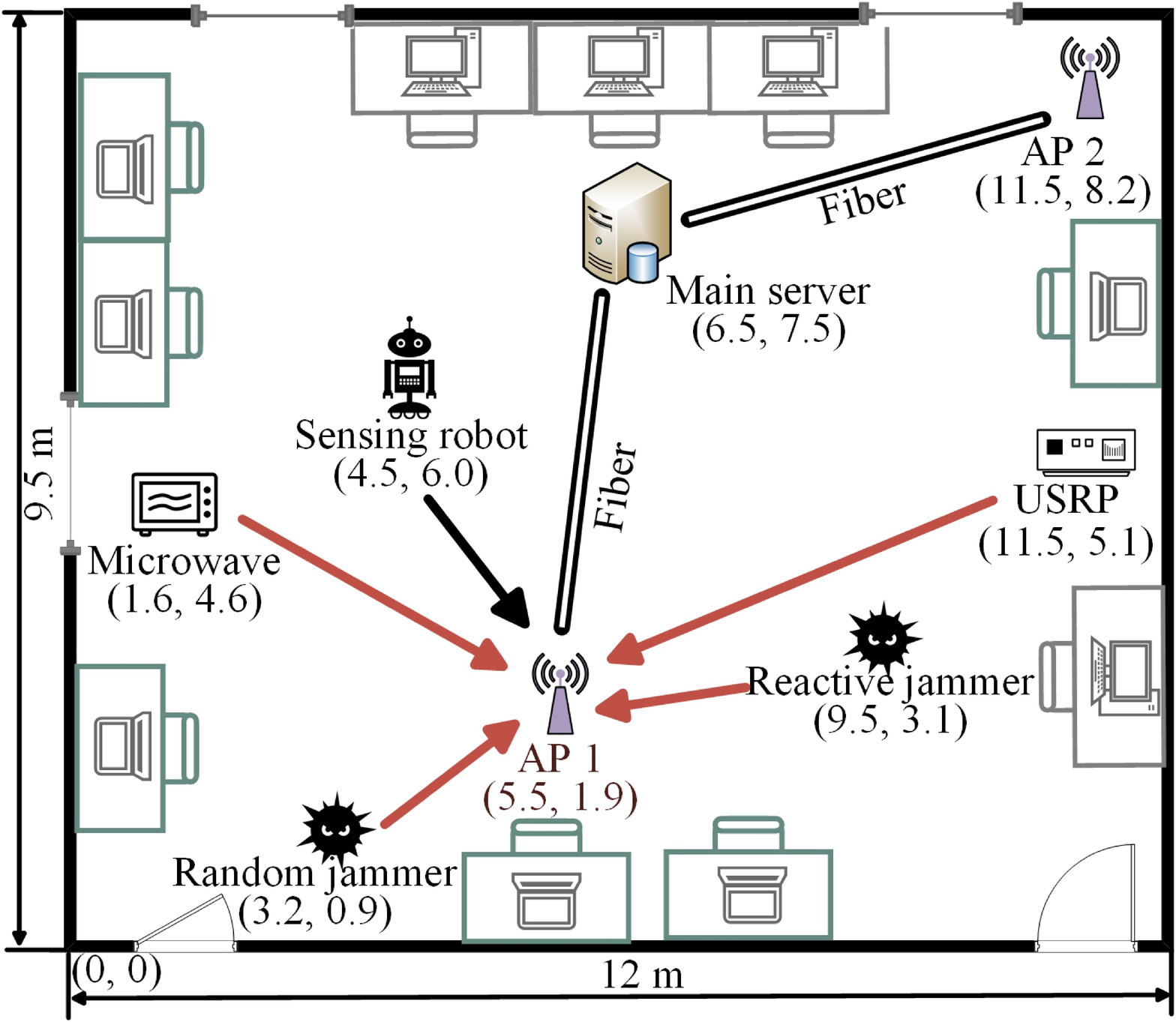}\\
\caption{Simulation topology in the sensing report collection of a sensing robot with two APs against a random jammer and a reactive jammer.}\label{EX2}
\end{center}
\end{figure}

In the second application, a mobile sensing robot collected the monitored data in the office and chose one of the $N$ channels to send sensing report to the main server via two APs against a random jammer located at (3.2m, 0.9m), a reactive jammer located in (9.5m,3.1m), and two interference sources, i.e., a microwave located at (1.6m, 4.6m) and a universal software radio peripherals system located at (11.5m, 5.1m), as shown in Fig. \ref{EX2}.

\begin{figure}[!t]\setlength{\abovecaptionskip}{0.cm}\setlength{\belowcaptionskip}{-0.5cm}
\begin{center}
\includegraphics[height=2.65 in]{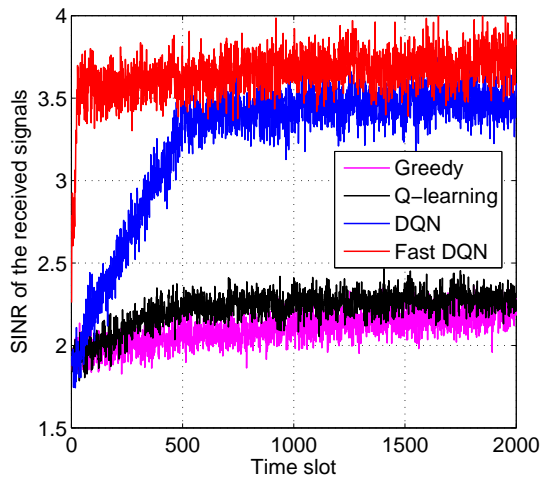}\\
{\footnotesize  (a) SINR of the mobile sensing robot signals}\\
\includegraphics[height=2.65 in]{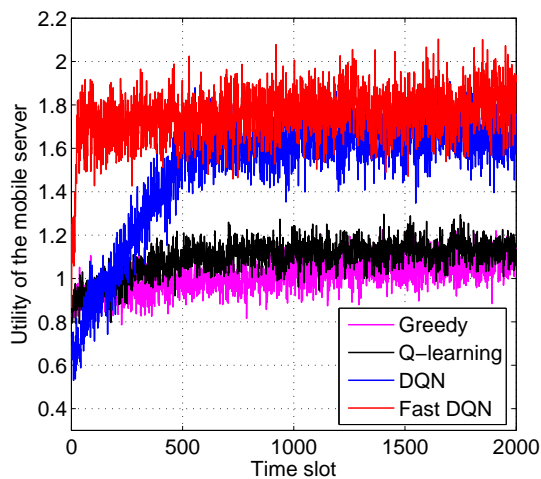}\\
{\footnotesize  (b) Utility of the mobile sensing robot}\\
\caption{Performance of the anti-jamming communication scheme in the sensing report transmission of a mobile sensing robot with 96 frequency channels in a dynamic game against a random jammer, a reactive jammer and two interference sources in the office as shown in Fig. \ref{EX2}.
}\label{SINR_Utility_time}
\end{center}
\end{figure}

\begin{figure}[!t]\setlength{\abovecaptionskip}{0.cm}\setlength{\belowcaptionskip}{-0.5cm}
\begin{center}
\includegraphics[height=2.65 in]{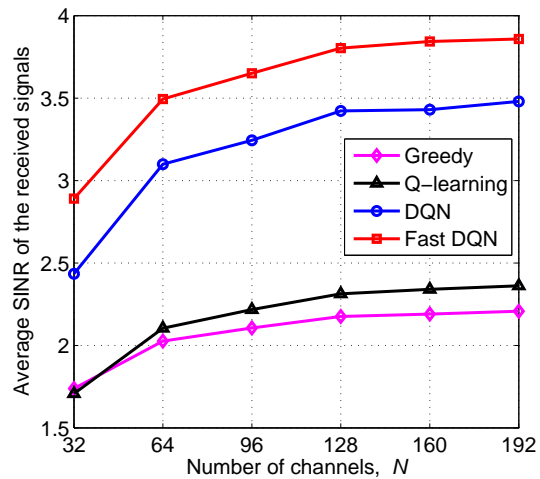}\\
{\footnotesize  (a) Average SINR of the mobile sensing robot signals}\\
\includegraphics[height=2.65 in]{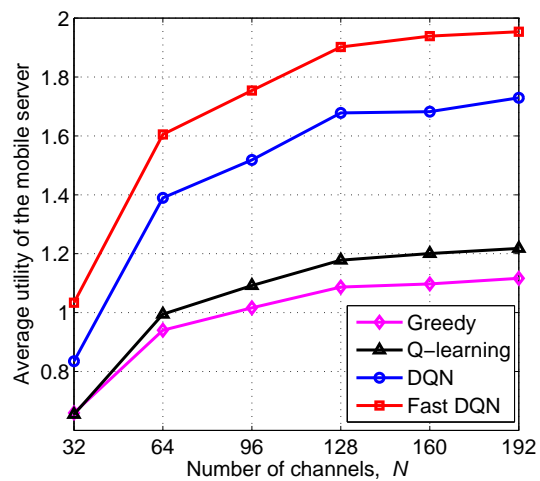}\\
{\footnotesize  (b) Average utility of the mobile sensing robot}\\
\caption{Average performance of the anti-jamming communication scheme in the sensing report transmission of a mobile sensing robot with $N$ frequency channels over 2000 time slots in each dynamic game and 200 scenarios against a random jammer, a reactive jammer and two interference sources in the office as shown in Fig. \ref{EX2}.
}\label{Avg_performance}
\end{center}
\end{figure}

In the simulation, the reactive jammer continuously monitors $N_r=8$ channels. The microwave interfered the serving AP at a probability 0.1 and the USRP system interfered the serving AP at a probability 0.05.

As shown in Fig. \ref{SINR_Utility_time}, the 2-D anti-jamming communication with the fast DQN-based scheme outperforms the DQN-based, the Q-learning based and the greedy-based schemes, with a faster learning speed, a higher SINR of the signals, and a higher utility. For instance, the fast DQN-based scheme converges after $50$ time slots, which saves $90$\% and $99.999$\% of the learning time compared with the DQN-based and the Q-learning based schemes, respectively. Therefore, the fast DQN-based scheme increases the SINR of the signals by $24.1$\% compared with the DQN-based scheme, which is $68.9$\% higher than that of the Q-learning based scheme at $300$-th time slot. Consequently, as shown in Fig. \ref{SINR_Utility_time}(b), the fast DQN-based scheme reaches the utility as high as $1.75$ which is $39.7$\% and $78.9$\% higher than that of the DQN-based and the Q-learning based schemes, respectively.

Fig. \ref{Avg_performance} shows that the proposed 2-D anti-jamming communication schemes can achieve higher SINR of the signals and higher utility of the mobile sensing robot with the number of channels increasing. For example, the average SINR of the signals with the fast DQN-based scheme increases by $31.8$\% to $3.81$, and achieves $84.1$\% higher average utility, if the number of channels increases from $32$ to $128$. The utility of the fast DQN-based scheme increases by $55.3$\% if the the number of channels increases from $32$ to $64$, and increases by $1.9$\% if the the number of channels increases from $128$ to $160$. In addition, the fast DQN-based scheme has the highest average SINR of the signals and the highest average utility in all of the four schemes. For instance, the fast-DQN based scheme achieves $12.8$\% higher SINR of the signals compared with the DQN-based scheme, which is $72.5$\% higher than that of the greedy-based scheme for the system with $64$ channels. Consequently, as shown in Fig. \ref{Avg_performance}(b), the average utility of the mobile sensing robot with the fast DQN-based scheme increases by $15.5$\% and $70.7$\% compared with the DQN-based and the greedy-based schemes, respectively.

\begin{figure}[!t]\setlength{\abovecaptionskip}{0.cm}\setlength{\belowcaptionskip}{-0.5cm}
\begin{center}
\includegraphics[height=2.65 in]{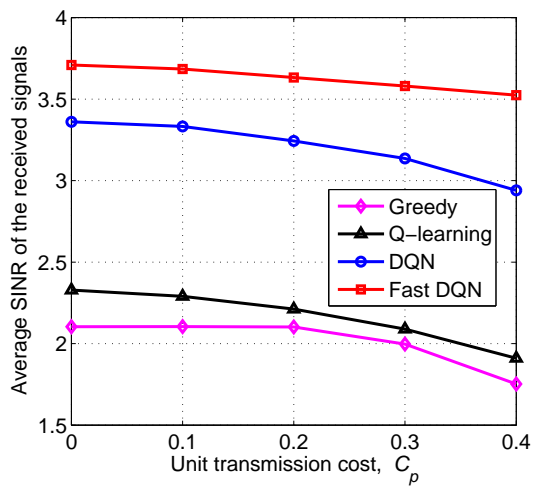}\\
{\footnotesize  (a) Average SINR of the mobile sensing robot signals}\\
\includegraphics[height=2.65 in]{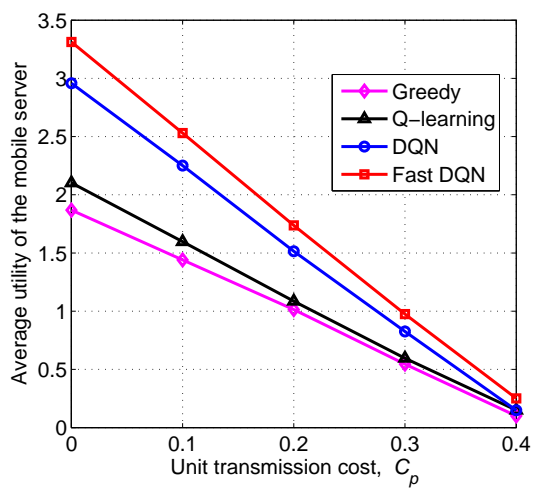}\\
{\footnotesize  (b) Average utility of the mobile sensing robot}\\
\caption{Average performance of the anti-jamming communication scheme in the sensing report transmission of a mobile sensing robot with 96 frequency channels over 2000 time slots in each dynamic game and 200 scenarios against a random jammer, a reactive jammer and two interference sources in the office as shown in Fig. \ref{EX2}.
}\label{Avg_deffhj}
\end{center}
\end{figure}

Fig. \ref{Avg_deffhj} illustrates the impacts of the unit transmission cost on the performance showing that both the average SINR of the signals and the average utility of the robot decreases with the unit transmission cost. For instance, the DQN-based scheme decreases the SINR of the signals by $4.9$\% and achieves $63.3$\% lower utility, if $C_p$ increases from $0.1$ to $0.3$. In addition, the anti-jamming performance of the DQN-based scheme exceeds that of the Q-learning based and the greedy-based schemes, and can be further improved by the fast DQN-based scheme. For example, the DQN-based scheme achieves $58.4$\% higher SINR of the signals and $56.3$\% higher utility than that of the greedy-based scheme, and be further increased by $16.7$\% and $19.4$\% with the fast DQN-based scheme, for the system with $C_p=0.1$.

\subsection{Sensing report collection against mobile jammers}
This simulation is based on the topology as shown in Fig. \ref{EX2} with two mobile jammers that change their locations randomly with probability 0.8 every 200 time slots.
\begin{figure}[!t]\setlength{\abovecaptionskip}{0.cm}\setlength{\belowcaptionskip}{-0.5cm}
\begin{center}
\includegraphics[height=2.65 in]{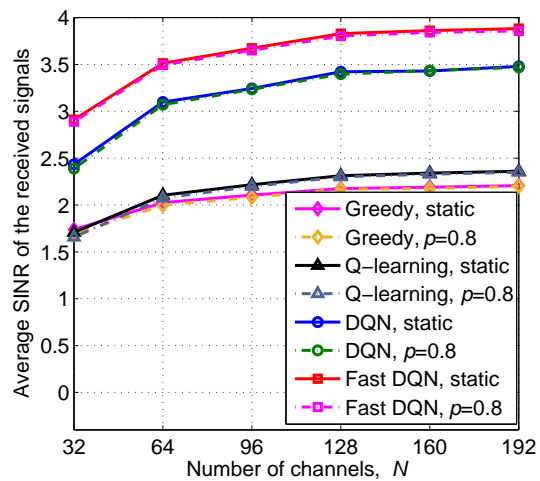}\\
{\footnotesize  (a) Average SINR of the mobile sensing robot signals}\\
\includegraphics[height=2.65 in]{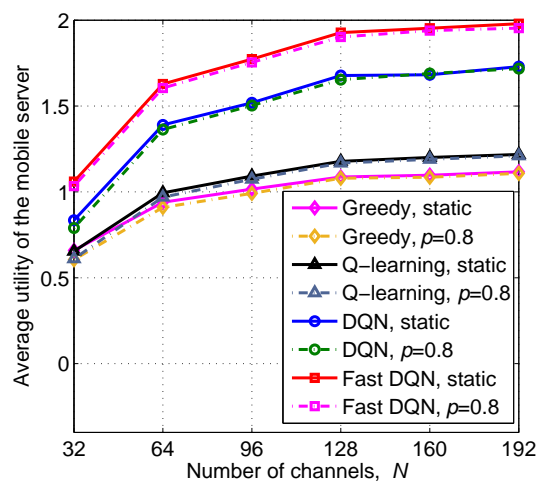}\\
{\footnotesize  (b) Average utility of the mobile sensing robot}\\
\caption{Average performance of the anti-jamming communication scheme in the sensing report transmission of a mobile sensing robot with $N$ frequency channels over 2000 time slots in each dynamic game and 200 scenarios against two  mobile jammers and two interference sources with $p=0.8$, in the office as shown in Fig. \ref{EX2}.
}\label{Avg_moveperformance}
\end{center}
\end{figure}

\begin{figure}[!t]\setlength{\abovecaptionskip}{0.cm}\setlength{\belowcaptionskip}{-0.5cm}
\begin{center}
\includegraphics[height=2.65 in]{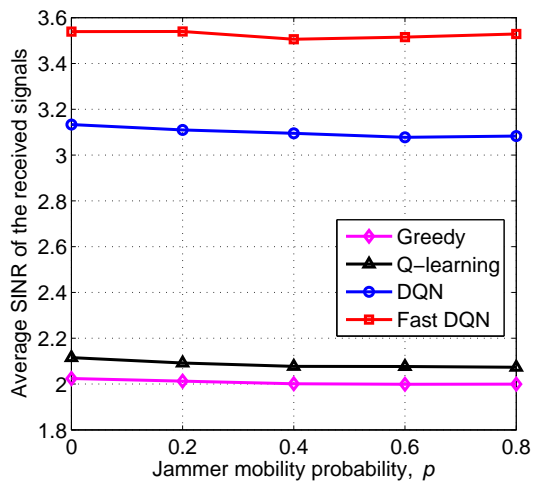}\\
{\footnotesize  (a) Average SINR of the mobile sensing robot signals}\\
\includegraphics[height=2.65 in]{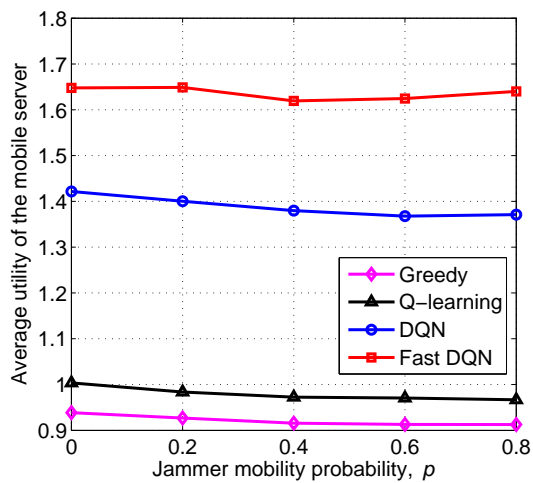}\\
{\footnotesize  (b) Average utility of the mobile sensing robot}\\
\caption{Average performance of the anti-jamming communication scheme in the sensing report transmission of a mobile sensing robot with $64$ frequency channels over 2000 time slots in each dynamic game and 200 scenarios against two mobile jammers and two interference sources, in the office as shown in Fig. \ref{EX2}.
}\label{Avg_deffmoveperformance}
\end{center}
\end{figure}

As shown Fig. \ref{Avg_moveperformance}, the proposed schemes are robust against the mobile jammers. For instance, the average SINR and the utility of the robot with the fast DQN-based scheme decrease by $0.6$\% and $1.1$\% if $N=96$ compared with the static jammers.

Fig. \ref{Avg_deffmoveperformance} illustrates the impacts of the jamming mobility, showing that the proposed schemes are robust against jamming mobility. For example, the SINR of the signals of the fast DQN-based scheme slightly decreases by $0.7$\% if the jammer mobility probability $p$ increases from $0$ to $0.6$ as shown in Fig. \ref{Avg_deffmoveperformance}(a). Consequently, as shown in Fig. \ref{Avg_deffmoveperformance}(b), the utility of the robot slightly decreases by $1.4$\% if $p$ increases from $0$ to $0.6$.

\section{Conclusions}\label{sec:conclusion}
In this paper, we have proposed an RL-based frequency-space anti-jamming mobile communication system that exploits spread spectrum and user mobility to resist cooperative jamming and strong interference. We have shown that, by applying a DQN-based frequency-space anti-jamming mobile communication scheme, a mobile device can achieve an optimal power allocation and moving policy, without being aware of the jamming and interference model and the radio channel model. Moreover, we have seen that the proposed fast DQN-based 2-D mobile communication scheme combines hotbooting, DQN and macro-actions can further accelerate the learning speed and thus improve the jamming resistance. Simulation results show that the fast DQN-based scheme increases the SINR of the signals compared with benchmark schemes. For instance, the fast DQN-based scheme saves $90$\% of the learning time required by DQN, and increases the SINR of the signals by $84.7$\% and increases $92.1$\% utility, compared with the greedy-based scheme.



\bibliography{my7}
\bibliographystyle{IEEEtr}
\begin{IEEEbiography}
[{\includegraphics[width=1in,height=1.2in,clip,keepaspectratio]{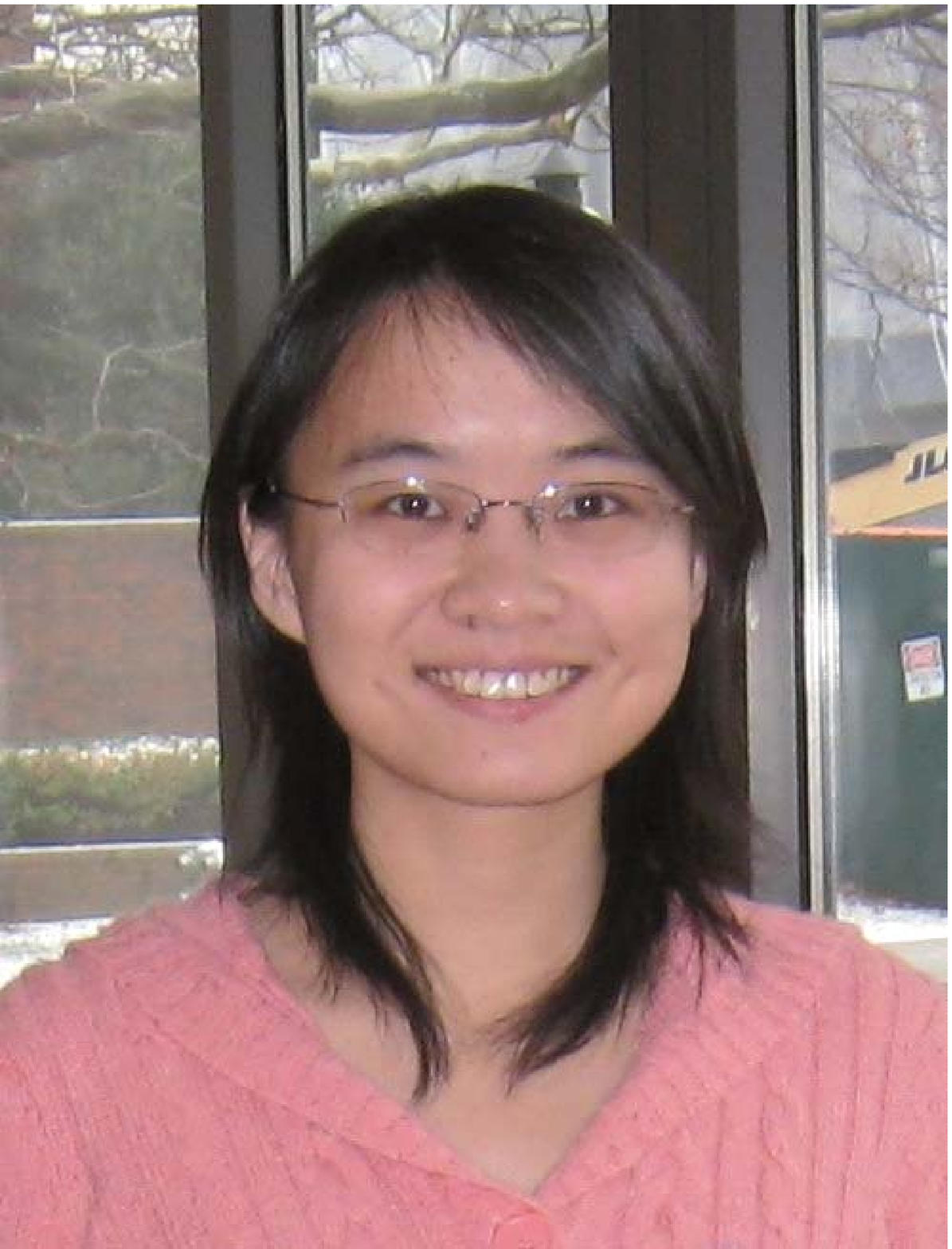}}]
{Liang Xiao}
is currently a Professor in the Department of Communication Engineering, Xiamen University, Fujian, China. She has served in several editorial roles, including an associate editor of IEEE Trans. Information Forensics \& Security and IET Communications. Her research interests include wireless security, smart grids, and wireless communications. She won the best paper award for 2016 IEEE INFOCOM Bigsecurity WS. She received the B.S. degree in communication engineering from Nanjing University of Posts and Telecommunications, China, in 2000, the M.S. degree in electrical engineering from Tsinghua University, China, in 2003, and the Ph.D. degree in electrical engineering from Rutgers University, NJ, in 2009. She was a visiting professor with Princeton University, Virginia Tech, and University of Maryland, College Park. She is a senior member of the IEEE.
\end{IEEEbiography}
\begin{IEEEbiography}
[{\includegraphics[width=1in,height=1.25in,clip,keepaspectratio]{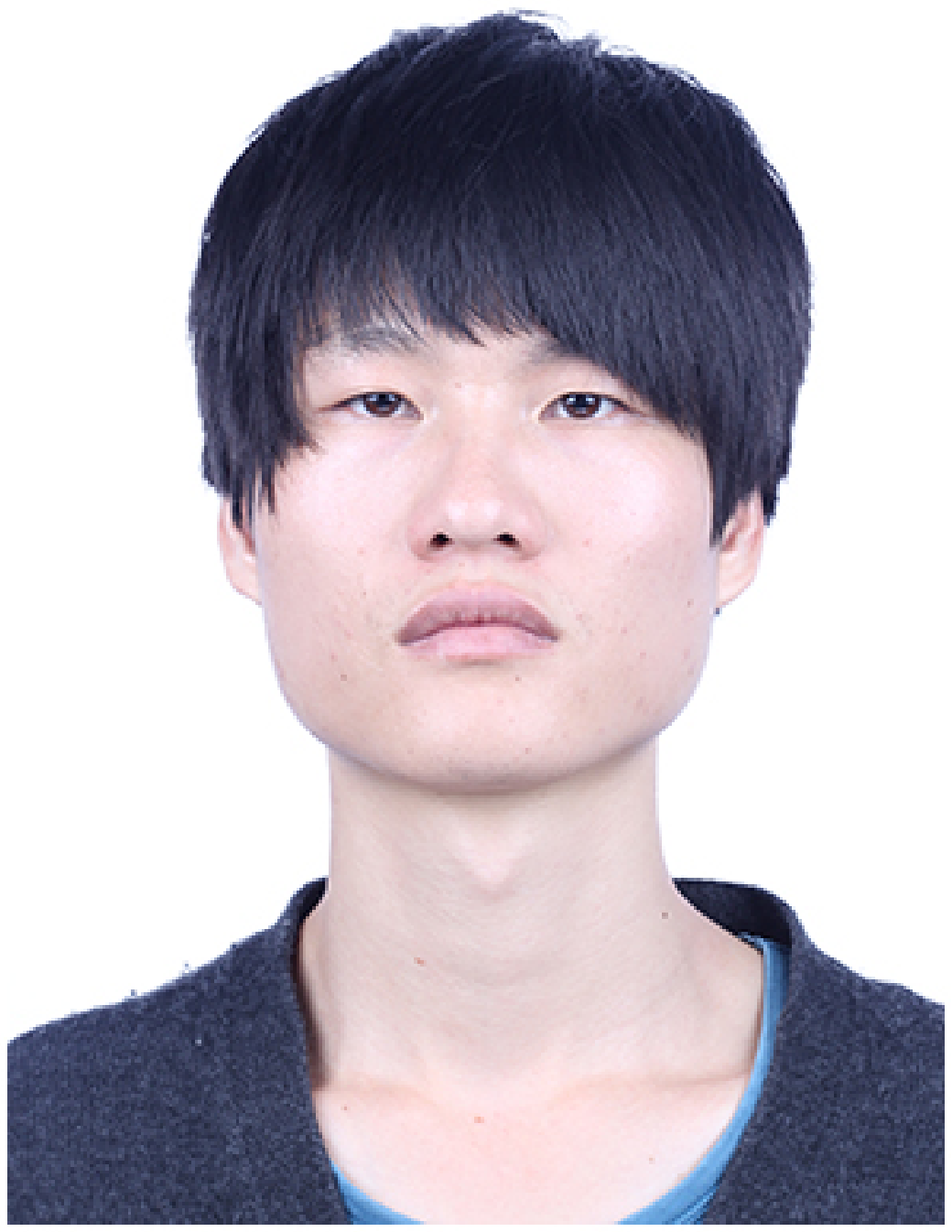}}]
{Guoan Han}
received the B.S. degree in communication engineering from Southwest Jiaotong University, Chengdu, China, in 2015, where he is currently pursuing the M.S. degree with the Department of Communication Engineering. His research interests include network security and wireless communications.
\end{IEEEbiography}
\begin{IEEEbiography}
[{\includegraphics[width=1in,height=1.25in,clip,keepaspectratio]{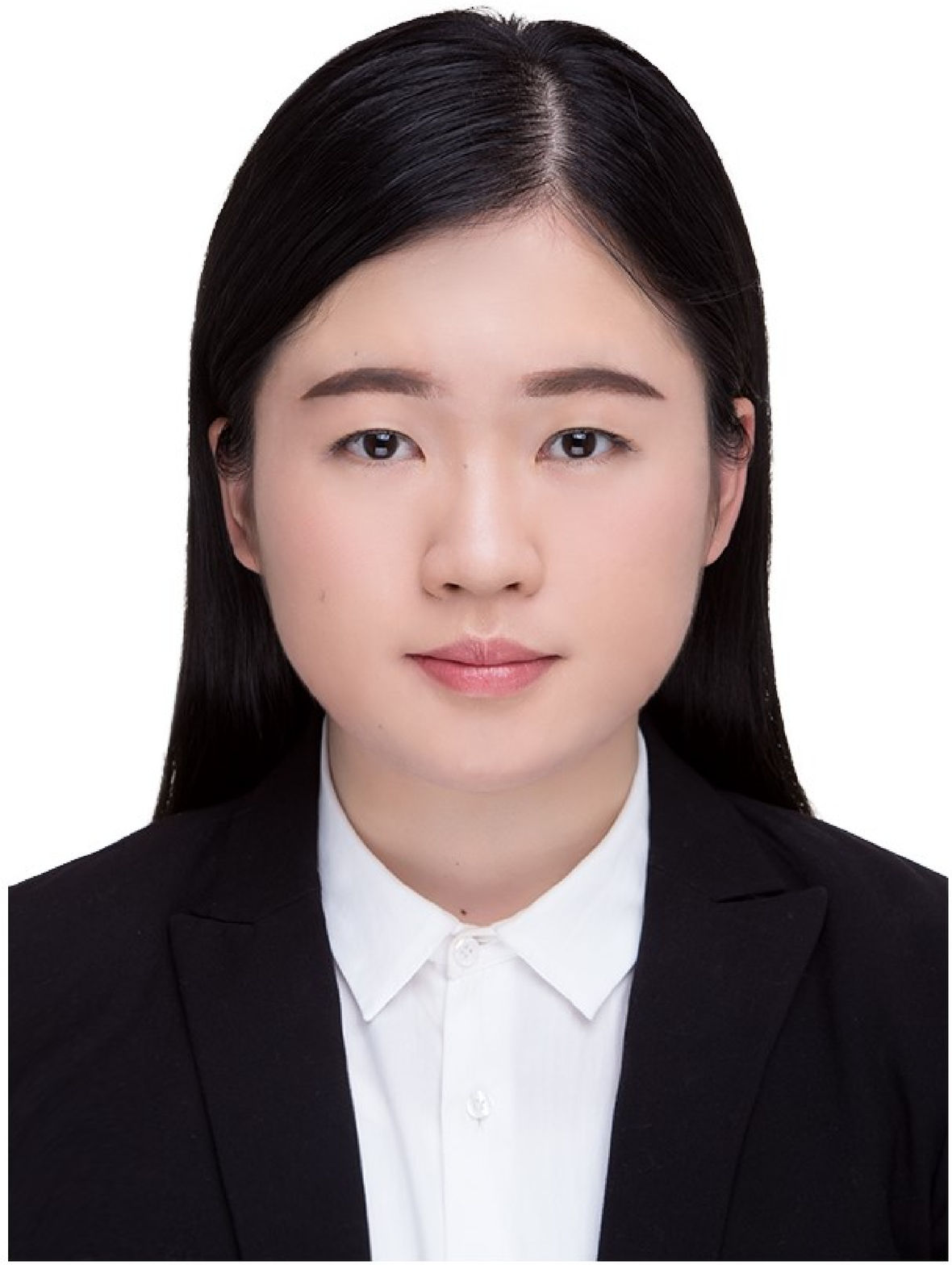}}]
{Donghua Jiang}
received the B.S. degree in electronic information science and technology from Xiamen University, Xiamen, China, in 2017, where she is currently pursuing the M.S. degree with the Department of Communication Engineering. Her research interests include network security and wireless communications.
\end{IEEEbiography}
\begin{IEEEbiography}
[{\includegraphics[width=1in,height=1.25in,clip,keepaspectratio]{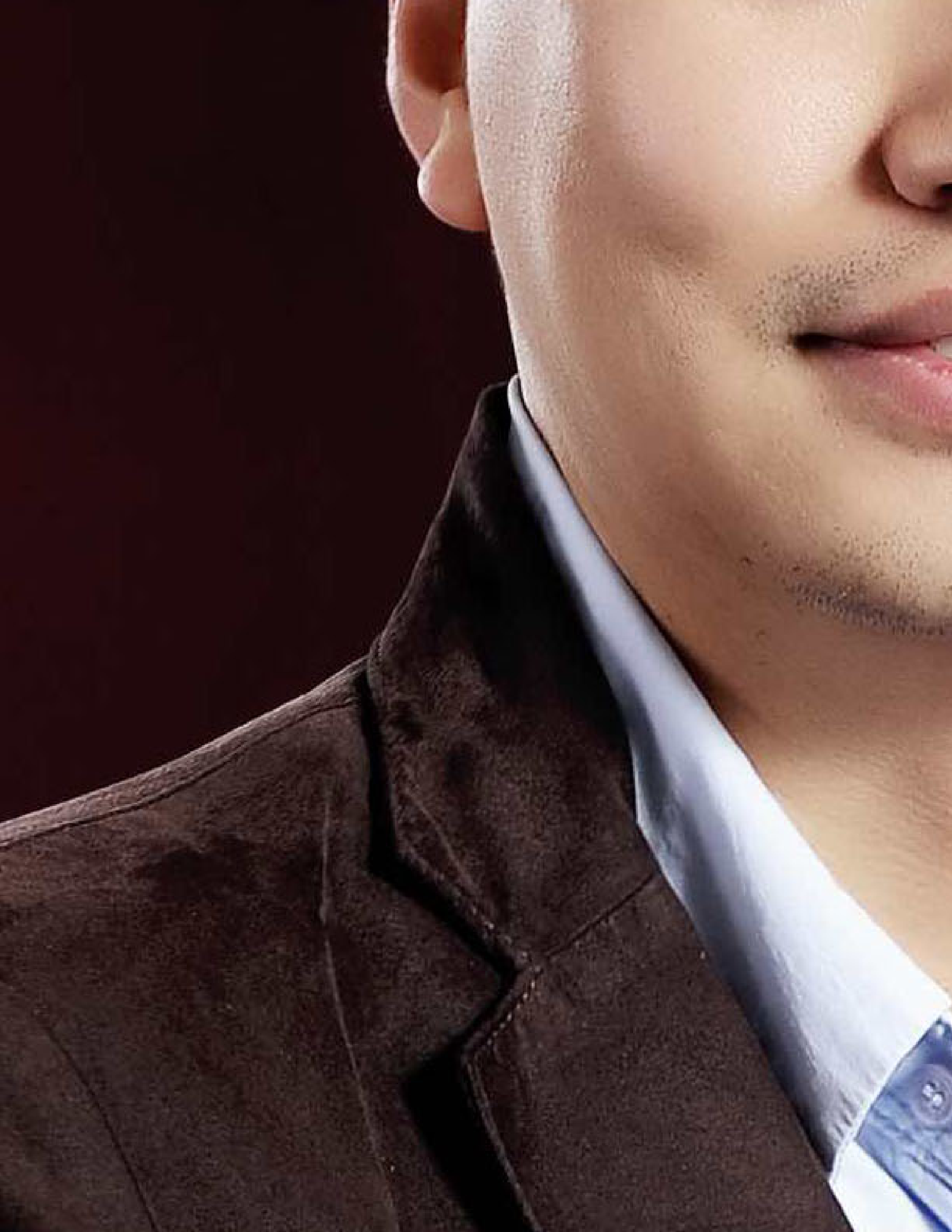}}]
{Hongzi Zhu}
Hongzi Zhu received his Ph.D. degree in computer science from Shanghai Jiao Tong University in 2009. He was a Post-doctoral Fellow in the Department of Computer Science and Engineering at Hong Kong University of Science and Technology and the Department of Electrical and Computer Engineering at University of Waterloo in 2009 and 2010, respectively. He is now an associate professor at the Department of Computer Science and Engineering in Shanghai Jiao Tong University. His research interests include vehicular networks, network and mobile computing. He received the Best Paper Award from IEEE Globecom 2016. He is a member of the IEEE Computer Society and Communication Society. For more information, please visit http://www.cs.sjtu.edu.cn/~hongzi/.
\end{IEEEbiography}
\begin{IEEEbiography}
[{\includegraphics[width=1in,height=1.25in,clip,keepaspectratio]{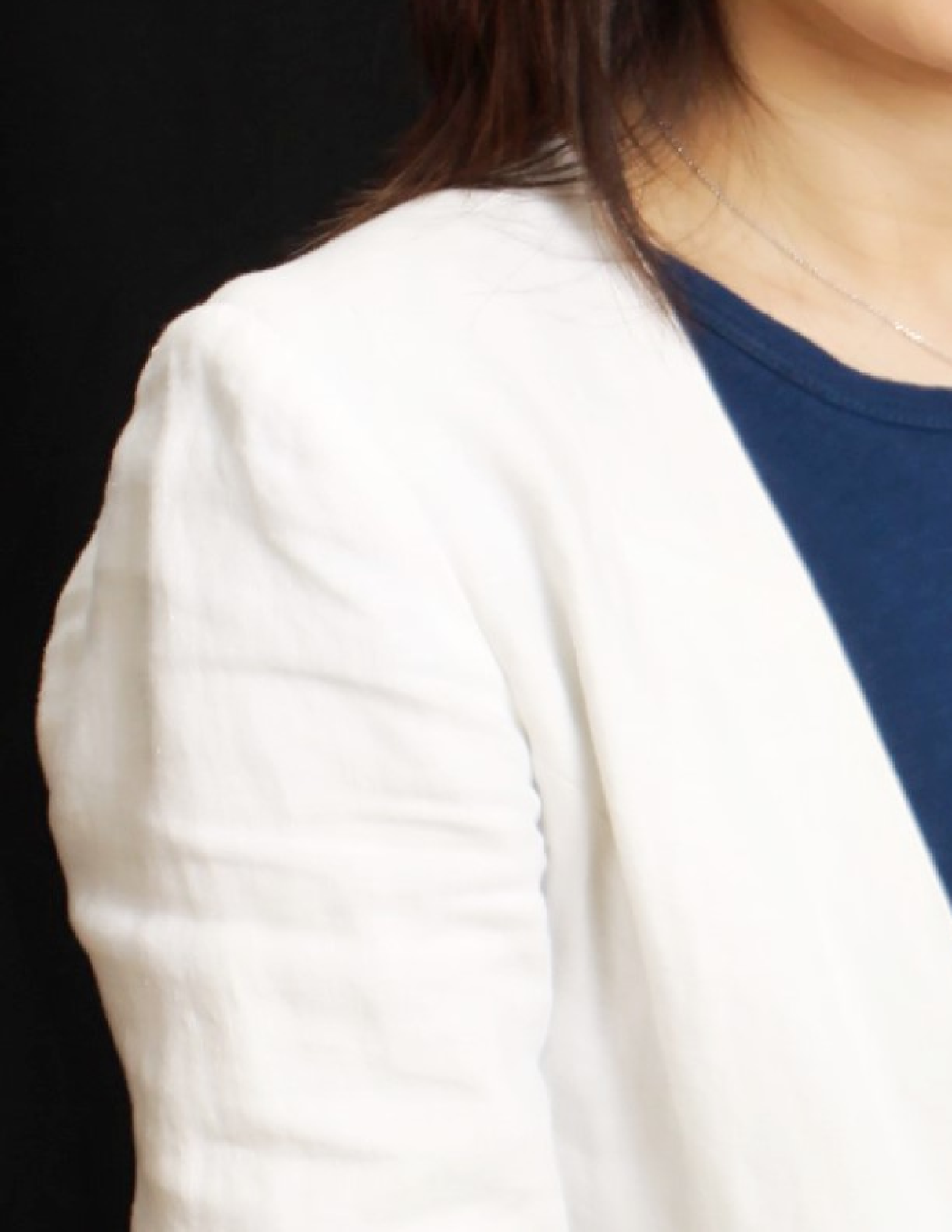}}]
{Yanyong Zhang}
has spent 15 yeras in the Electrical and Computer Engineering Department at Rutgers University, where she is currently a Professor. She is also a member of the Wireless Information Networks Laboratory (Winlab). She has 20 years of research experience in the areas of sensor networks, mobile computing and high-performance computing, and has published more than 100 technical papers in these fields. She is an Associate Editor for IEEE TMC, IEEE TSC, IEEE/ACM ToN, and Elsevier Smart Health. She is a recipient of NSF CAREER award. She received her Ph.D. from Penn State in 2002.
\end{IEEEbiography}
\begin{IEEEbiography}
[{\includegraphics[width=1in,height=1.25in,clip,keepaspectratio]{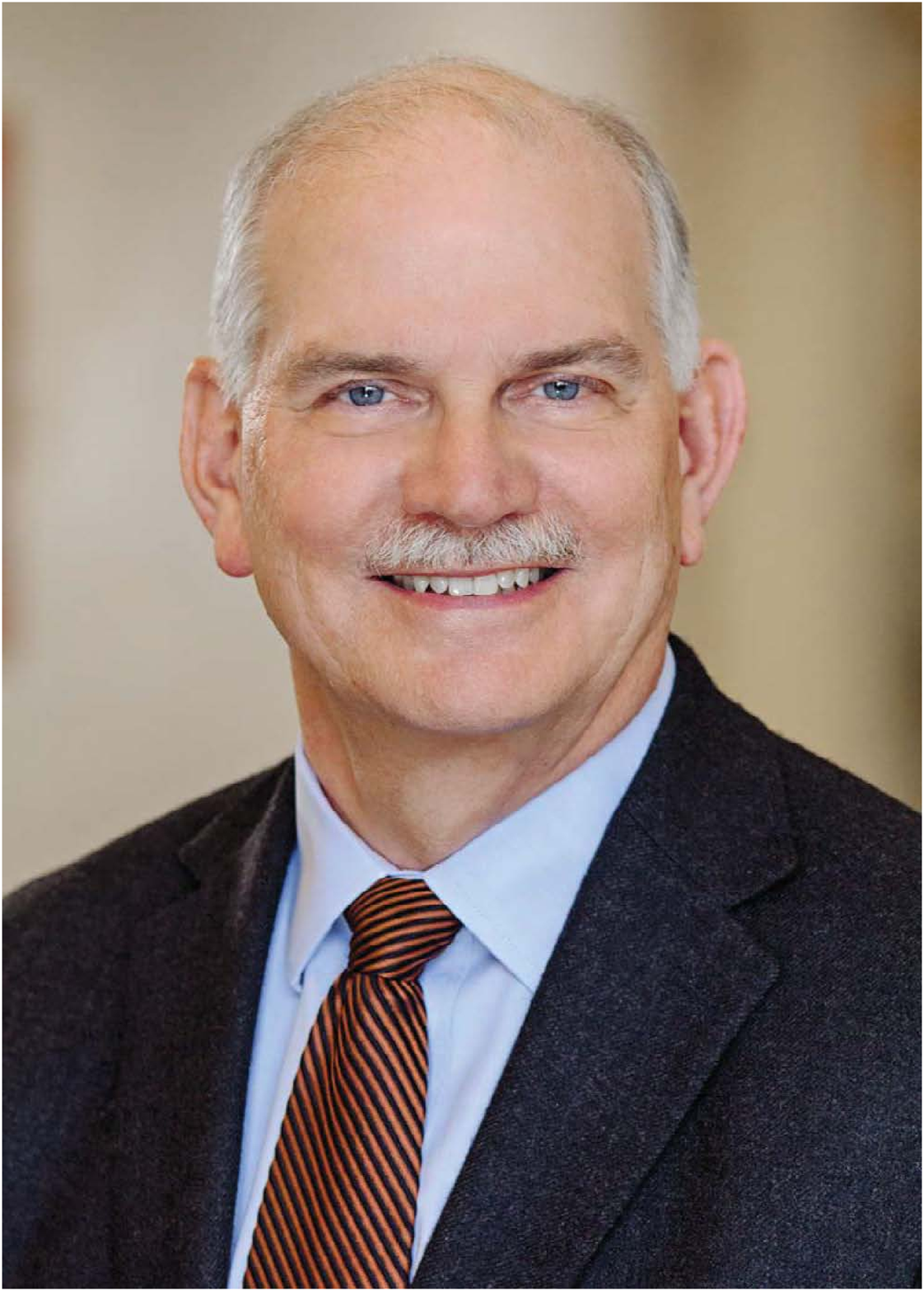}}]
{H. Vincent Poor}
(S'72, M'77, SM'82, F'87) received the Ph.D. degree in EECS from Princeton University in 1977.  From 1977 until 1990, he was on the faculty of the University of Illinois at Urbana-Champaign. Since 1990 he has been on the faculty at Princeton, where he is currently the Michael Henry Strater University Professor of Electrical Engineering. During 2006 to 2016, he served as Dean of Princeton’s School of Engineering and Applied Science. His research interests are in the areas of information theory, statistical signal processing and stochastic analysis, and their applications in wireless networks and related fields. Among his publications in these areas is the book Mechanisms and Games for Dynamic Spectrum Allocation (Cambridge University Press, 2014).

Dr. Poor is a member of the National Academy of Engineering, the National Academy of Sciences, and is a foreign member of the Royal Society. He is also a fellow of the American Academy of Arts and Sciences, the National Academy of Inventors, and other national and international academies. He received the Marconi and Armstrong Awards of the IEEE Communications Society in 2007 and 2009, respectively. Recent recognition of his work includes the 2016 John Fritz Medal, the 2017 IEEE Alexander Graham Bell Medal, a Doctor of Science honoris causa from Syracuse University (2017) and Honorary Professorships at Peking University and Tsinghua University, both conferred in 2016.
\end{IEEEbiography}

\end{document}